\documentclass[letterpaper,twocolumn,10pt]{article}
\usepackage{usenix2019_v3}

\usepackage{tikz}
\usepackage{amsmath}
\usepackage{color,soul}

\usepackage{filecontents}

\usepackage{cite}
\usepackage{amsmath,amssymb,amsfonts}
\usepackage{algorithmic}
\usepackage{graphicx}
\usepackage{textcomp}
\usepackage{soul,color}
\usepackage{xcolor}
\usepackage{multirow}
\usepackage{subcaption}
\usepackage{placeins}
\usepackage{chngcntr}
\usepackage{authblk}

\begin{document}
%-------------------------------------------------------------------------------

%don't want date printed
\date{}

\title{\Large \textbf{PINCH: An Adversarial Extraction Attack Framework for Deep Learning Models\\}}

%\author[1]{Anonymous author(s)}
\author[1]{William Hackett}
\author[1]{Stefan Trawicki}
\author[1]{Zhengxin Yu}
\author[1,2]{Neeraj Suri}
\author[1,2]{Peter Garraghan}

\affil[1]{Lancaster University}
\affil[2]{Mindgard}

\maketitle

%-------------------------------------------------------------------------------
\begin{abstract}

Adversarial extraction attacks constitute an insidious threat against Deep Learning (DL) models in-which an adversary aims to steal the architecture, parameters, and hyper-parameters of a targeted DL model. Existing extraction attack literature have observed varying levels of attack success for different DL models and datasets, yet the underlying cause(s) behind their susceptibility often remain unclear, and would help facilitate creating secure DL systems. In this paper we present \textit{PINCH}: an efficient and automated extraction attack framework capable of designing, deploying, and analyzing extraction attack scenarios across heterogeneous hardware platforms. Using PINCH, we perform extensive experimental evaluation of extraction attacks against 21 model architectures to explore new extraction attack scenarios and further attack staging. Our findings show (1) \textit{key extraction characteristics} whereby particular model configurations exhibit strong resilience against specific attacks, (2) even partial extraction success enables further staging for other adversarial attacks, and (3) equivalent stolen models uncover differences in expressive power, yet exhibit similar captured knowledge.

\end{abstract}

%Ascertaining such root-cause weaknesses would help facilitate secure DL models, though due to the overwhelmingly high technical effort and time required to implement and evaluate an attack makes it infeasible to explore the large number of unique extraction attack scenarios.

\section{Introduction} \label{sec:intro}
%-------------------------------------------------------------------------------

Deep Learning (DL) has become a critical technology supporting a growing diversity of applications.
However, the successful deployment and execution of DL models is threatened by cyber attacks occurring within systems \cite{stealingPredictionAPIs, duddu2019stealing, deepSniffer, mlAttacksSurvey}, compromising DL model integrity, privacy, and confidentiality~\cite{mlSecuritySurvey}. 
A particularly damaging threat against DL models are \textit{extraction attacks} (also known as \textit{model stealing}). Extraction attacks occur when an adversary attempts to extract fundamental characteristics of a target DL model (architecture, parameters, hyper-parameters)~\cite{zhu2020hermes, LeakyDNNSideChannel} to reconstruct an identical or highly similar DL model~\cite{jagielski2020high}. Such attacks result in information leakage, digital IP theft, and enable further DL model attacks to be staged \cite{zhu2020hermes, deepSniffer, surveyWANG201912}.

Extensive studies of DL model extraction attacks have been conducted to understand and mitigate their impact~\cite{mlSecuritySurvey, activeLearningExtraction}. However these studies have predominately been performed with isolated attacks, each leveraging distinctive threat models and deployment scenarios with different DL model types, datasets, and hardware platforms. Given that extraction attacks yield varying degrees of success when exposed to different DL model types and datasets~\cite{LeakyDNNSideChannel, stealingPredictionAPIs, yuan2020es, deepSniffer}, it is necessary to study extraction attacks across a multitude of deployment scenarios to determine whether there exist common associations between extraction attack success, DL model characteristics and platform hardware properties.
% Therefore there is a need to study whether there are associations between extraction attack success, generic DL model characteristics, and platform types (GPU architecture, CPU AVX instruction availability). 

% Overall, existing inference attacks have been studied under different threat models and experimental settings, albeit
% in isolation. This prompts the need for a holistic understanding of the risks caused by these attacks, such as the scenarios different inference attacks can be applied to, the common
% factors that influence these attacks’ performance, and the relations among the attacks, as well as the overall effectiveness
% of defense mechanisms.

% Yet the reasoning behind such phenomena remains unexplained, and precisely what characteristics are key in determining attack success when considering the wide variety of models and platforms remains entirely untested.

Attaining such knowledge is constrained by the overwhelmingly high technical effort (and time) required to understand, implement and evaluate the large number of unique extraction attacks, platforms and DL model architectures in existence. This is because current studies are bespokely designed to operate for a targeted or small sub-set of deployment scenarios (e.g. a single hardware platform or DL model architecture) \cite{deepRecon, deepSniffer, stealingPredictionAPIs}. Whilst this approach is effective to demonstrate extraction attack feasibility, it is not possible to study attack effectiveness and generalizability without extensive re-designing and engineering of attacks to operate within different operational scenarios \cite{convnextArchs, transformers, nvidiaTransformerAccelerator}.

%ML Doctor: Framework only focuses on inference attacks (We additionally look at other system level attacks; DeepSniffer etc.)
%PG - Targeting only one stack and model characteristics results in a limitation in studying common/generalizable features of extraction attacks (described a couple paragraphs above.
Extensive progress has been made to create extraction attack frameworks to alleviate the complexity of re-implementing attacks and providing configurable attack scenarios \cite{Nicolae2018AdversarialRT, PrivacyRaven, mldoctor}. However, such frameworks exhibit limitations towards studying generalizable features of extraction attacks, as current proposed frameworks provide discrete approaches towards extraction, typically only implementing attacks within one area of the DL system attack surface and targeting specific model characteristics \cite{mldoctor, zhu2020hermes}. Additionally, current frameworks are often limited to executing smaller (often bespoke) models and datasets, unable to evaluate larger models and complex dataset pairings deployed within the modern DL landscape.

%This highlights an increasing need for a generalised framework that enables the execution of various extraction attacks targeting a more of the DL system stack and model characteristics. 
%\todo{Z: The last sentence in this para is overlapped with the before. I removed it. Also this para can be combined with the previous para and could put it in front} 

%In this paper we present PINCH: an efficient and automated extraction attack framework capable of deploying and evaluating multiple DL models and attacks across heterogeneous hardware platforms.

To tackle limitations of existing work, we present PINCH: an efficient and automated extraction attack framework capable of deploying a large number of DL models, attacks, and deployment environments in a generalizable manner. Our framework performs (1) \textit{dynamic framework-independent model loading and training} via transfer learning and curated AI deployment repositories, with (2) configuration of attacks encapsulated as \textit{attack scenarios}, and (3) \textit{experiment automation} for recording and reporting.

% \hl{\textbf{ST- maybe re-state these, online repositories doesn't sound the best}}

PINCH is capable of automated attack execution that extracts DL model characteristics utilizing multiple areas of the DL system attack surface, enabling exploration of scenarios not examined within contemporary literature, and provides support for unexplored adversarial attack staging. We demonstrate the effectiveness of PINCH by empirically evaluating extraction scenarios across different state-of-the-art extraction attack types \cite{knockOffDLs,deepSniffer,deepRecon} when exposed to various DL model architectures, datasets, and hardware/software environments. Our work makes the following contributions:

 \begin{itemize}
     \item \textbf{PINCH}: An end-to-end automated adversarial attack framework capable of efficiently executing extraction attack scenarios and enabling detailed evaluation across a variety of model families, architectures, datasets, and hardware platforms (Section  \ref{sec:frameworkdesign}).
     \item \textbf{Key extraction characteristics}: Through extensive experimentation of hundreds of unique extraction attack scenarios, we have identified key extraction attack characteristics that affect success spanning model architecture, and dataset complexity, and hardware (Section \ref{sec:evaluation}).
     \item \textbf{Further attack staging}: We demonstrate the feasibility of adversarial attack staging. Discovering it is possible to launch successful model inversion attacks on DL models created from partially successful extraction attacks, additionally uncovering limitations in existing methods for measuring DL model similarity for denoting attack success (Section \ref{sec::furtherAttacks}).
     %\item \textbf{Secure AI insights}. We have identified several new phenomena in adversarial attacks. (1) The existence of \textit{intrinsic resistance} whereby specific model configurations exhibited strong resilience to specific attacks (Section  \ref{sec:intrinsicResistance}), and (2) Stolen models can exhibit equivalent target model performance, yet can be composed of uniquely different DL model characteristics (Section  \ref{sec:equivalency}).
     \item \textbf{Stolen model equivalency}: We have identified that stolen models can exhibit equivalent target model performance, yet exhibit similar captured knowledge while being composed of uniquely different DL model characteristics (Section \ref{sec:equivalency}).
 \end{itemize}

% The paper is structured as follows: Section \ref{sec:background} introduces the background of DL systems, model extraction, and discusses current challenges in extraction frameworks. Section \ref{sec:threatmodel} presents the threat model. Section \ref{sec:extractionattack} discusses the extraction attacks involved in our study. Section \ref{sec:frameworkdesign} outlines the component design and implementation of our framework. The experiment setup is described in Section \ref{sec:setup}. Section \ref{sec:evaluation} conducts an empirical evaluation of extraction attacks within DL systems. Section \ref{sec:discussion} discusses analysis findings. Section \ref{sec:relatedwork} reviews related work, and Section \ref{sec:conclusion} concludes the paper.

%-------------------------------------------------------------------------------
\section{Background} \label{sec:background}
%-------------------------------------------------------------------------------

\subsection{Deep Learning Systems}
\textit{Deep Learning (DL)} is a sub-field of \textit{Machine Learning (ML)}, which uses multiple processing layers to learn representations from input data with multiple levels of abstraction \cite{lecun2015deep}. \textit{DL models} are represented by \textit{Deep Neural Networks} (DNNs); collections of \textit{Operators} (Convs, MaxPool, ReLU, etc.), specialized programs designed for performing actions on tensors, grouped into \textit{Layers}. A DNNs operator layers are selected and organized based on desired \textit{architecture} best suited for different applications, \textit{e.g.}, Convolutional Neural Networks (CNNs) for image classification, and Long Short-Term Memory (LSTM) for time-series data analysis. DL models leverage accelerator devices such as \textit{Graphics Processing Units} (GPUs) that enable parallel execution of operators, hastening the training process\cite{facebooktraining}, as well as performing faster model inference. A \textit{Deep Learning System (DL system)} provides, CPU, accelerators and the accompanying software (ML frameworks, libraries) to perform DL model training and inference.

%A machine equipped with a CPU, accelerators and the accompanying software (ML frameworks, libraries) to perform DL model training and inference is a \textit{Deep Learning System (DL system)}.

%---------------------------
\begin{figure}[t]
\begin{center}
\includegraphics[width=\linewidth]{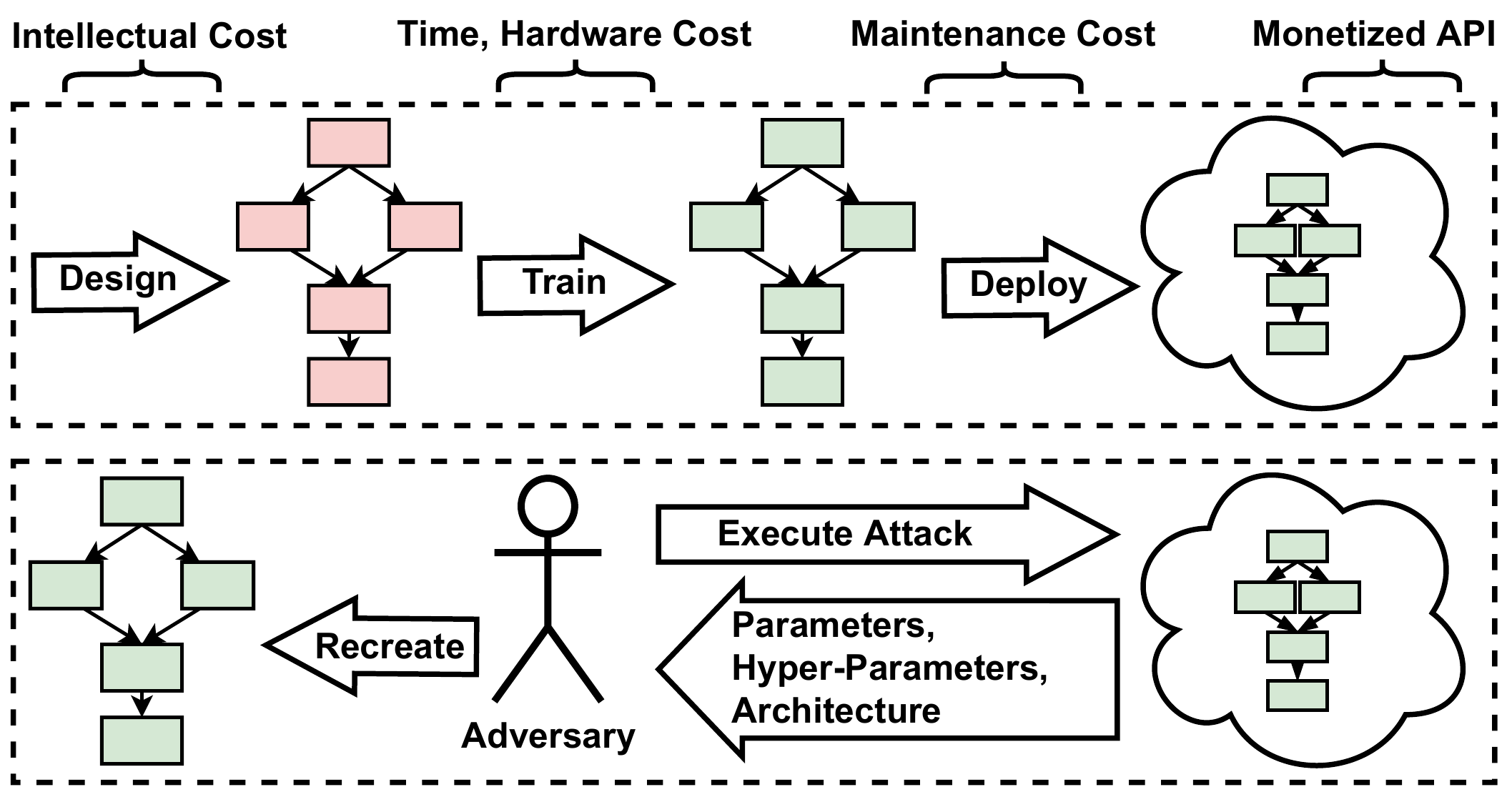}
\end{center}
\caption{\label{fig:adversary} \textbf{Overview of extraction attack process}. Deep Learning models comprising of architecture, parameters and hyper-parameters can be be stolen via extraction attacks.}
\end{figure}
%% %---------------------------

DL systems have been widely adopted throughout both industry and research, providing considerable acceleration to the creation of cutting-edge DL models capable of performing tasks unknown to previous generational systems \cite{gpuAcceleratorSurvey}. The widespread usage of such systems has lead to increasing concern of privacy and security related issues surrounding deployed DL models. The increased data and sophistication present in DL models has made DL systems a target for \textit{adversarial attacks}, aiming to perform attacks spanning model evasion \cite{evasionAttacks}, poisoning \cite{poisoningAttack}, as well as extract sensitive and confidential data \cite{mlAttacksSurvey, surveyWANG201912, zhu2020hermes}. The concern raised from the existence of such attacks has prompted extensive research to understand adversarial attacks \cite{AdversarialMachineLearning, surveyWANG201912, mlAttacksSurvey}, and protection against successful attack execution \cite{resnetPaper, mlSecuritySurvey}.

\subsection{Model Extraction}

\textit{Model extraction}, also referred to as \textit{model stealing}, is a set of adversarial attacks that extract fundamental \textit{characteristics} of a DL model: its architecture, parameters and hyper-parameters (Figure \ref{fig:adversary}). A \textit{stolen model} is created using extraction techniques to collect information leakage (model characteristics) via access to a target DL model or its underlying DL system, and \textit{recreating} a copy of the \textit{target model} \cite{zhu2020hermes, deepSniffer, LeakyDNNSideChannel}. A stolen model can be used for further attacks \cite{stealingPredictionAPIs} or designing a replica model with similar performance \cite{zhu2020hermes, stealingPredictionAPIs}.

%\textbf{Extraction techniques}. DL models can be stolen across a wide range of attack surfaces covering various areas of the DL system attack surface. For instance, an adversary can perform prediction API attacks by obtaining predictions on input feature vectors to train a local substitute model \cite{stealingPredictionAPIs, cloudLeak, yuan2020es}, or a side-channel attack by extracting information leakage from Peripheral Component Interconnect Express (PCIe) traffic. Many works have demonstrated that system operation (\textit{i.e.}, timing, power usage, computation, cache) can be exploited to infer the underlying operators of the DL model, which can be exploited in order to perform model extraction \cite{LeakyDNNSideChannel, deepSniffer, deepRecon}.  
%Additionally, attacks have different numbers of intermediate stages depending on their complexity, based on the \textit{MITRE ATLAS (MA)} \cite{atlas} knowledge base. PINCH focuses on 4 MA \textit{tactics} and their enabling \textit{techniques}: 1) \textit{Initial Access}, where the adversary prepares the environment such as deploying spy kernels and monitoring code \cite{zhu2020hermes, deepSniffer, deepSniffer}. 2) \textit{Attack Staging}, wherein preliminary attacks are launched to gather DL model system and model information \cite{deepSniffer, deepSniffer}. 3) \textit{Exfiltration}, primarily the deployment of API attacks, potentially using previously gathered information from attack staging. \cite{LeakyDNNSideChannel, RenderedInsecure, zhu2020hermes}.

\textbf{Extraction techniques}. DL models can be stolen across a wide range of attack surfaces covering various areas of the DL system stack. For instance, an adversary can perform prediction API attacks by obtaining predictions on input feature vectors to train a local substitute model \cite{stealingPredictionAPIs, cloudLeak, yuan2020es}, or a side-channel attack by extracting information leakage from Peripheral Component Interconnect Express (PCIe) traffic. Many works have demonstrated that system operation (\textit{i.e.}, timing, power usage, computation, cache) can be exploited to infer the underlying operators of the DL model and used to perform model extraction \cite{LeakyDNNSideChannel, deepSniffer, deepRecon}. Based on \textit{MITRE ATLAS (MA)}, attacks have different numbers of intermediate stages depending on their \textit{tactics} and their enabling \textit{techniques} \cite{atlas}. (1) \textit{Initial Access}, where the adversary prepares the environment such as deploying spy kernels and monitoring code \cite{zhu2020hermes, deepSniffer}. (2) \textit{Attack Staging}, wherein preliminary attacks are launched to gather DL model system and model information \cite{deepSniffer}. (3) \textit{Exfiltration}, primarily the deployment of API attacks, potentially using previously gathered information from attack staging. \cite{LeakyDNNSideChannel, RenderedInsecure, zhu2020hermes}.

\textbf{Model recreation}. Recreation focuses on training a model, either uninitialized or pre-trained, that provides an architecture and weights \cite{cloudLeak, deepSniffer} with the intention of replicating a target model by leveraging collected information leakage of DL model characteristics. Extracted target model characteristics can be acquired using a number of adversarial attacks, such as a training set created from synthetic or ground truth inputs paired with confidence values or labels gained from a prediction API attack \cite{stealingPredictionAPIs, cloudLeak, zhu2020hermes}. It is feasible for sophisticated approaches using side-channels to observe DL system traffic to infer DL model characteristics \cite{zhu2020hermes, deepSniffer, LeakyDNNSideChannel}. Using gathered metrics, GPU kernels mappings and inference inputs and outputs enable the creation of a dataflow graph representing model architecture layout. The complexity of model recreation can be vast due to the possible combinations of ML frameworks, compute libraries, model architectures, among other variables that can be partially and entirely unknown to an adversary when attacking a DL system. 

\textbf{Consequence}. Failure to defend against extraction attacks can compromise the integrity, privacy and confidentiality of the DL system. System integrity can be compromised during attack preparation and execution, with the potential backdoors created for future access \cite{mlModelBackDoor}. Data privacy is degraded via stolen model characteristics being exploited to stage further attacks that extract training data information \cite{stealingPredictionAPIs, modelInversion}. Furthermore, the confidentiality of the DL model is compromised since adversaries have access to model characteristics, therefore allowing adversaries to reverse engineer and steal confidential data, such as the training dataset.

\subsection{Challenges in Extraction Attack Research} \label{2C}

% W - 'Deployment Scenarios' assuming this refers to the DL environment. Highlighted due to usage of the word scenarios. Overlap with extraction scenarios?
In model extraction literature, a number of studies have demonstrated the feasibility and practicality of extraction attacks against DL models \cite{knockOffDLs, zhu2020hermes, stealingPredictionAPIs}. However these studies are predominately performed with isolated attacks targeting one DL model characteristic, leveraging distinctive threat models, and bespoke deployment scenarios with different DL types \cite{LeakyDNNSideChannel, stealingPredictionAPIs, yuan2020es, deepSniffer}. This is problematic given the evolving DL landscape of extraction scenarios whereby current extraction attack implementations are obsolete when paired with state-of-the-art DL types \cite{convnextArchs, transformers}, and DL systems \cite{nvidiaTransformerAccelerator, v100GPU}. 

% W - This paragraph should highlight the existence of common associations within extraction attack literature, why it's difficult to explore due to limitations, and finally why fixing this limitation will benefit extraction literature.
From extraction attack literature, it is observable that extraction attacks yield varying degrees of success when exposed to different DL model types and datasets \cite{stealingPredictionAPIs, zhu2020hermes}. However, current extraction attacks lack the generalizability required to execute attacks across different extraction scenarios. Exploring common associations is challenging due to the technical effort required to conduct attacks across attack scenarios in software and hardware heterogeneous DL systems. Understanding the associations attributed to DL types, datasets, and deployment scenarios can greatly benefit the fundamental understanding towards varying extraction success observed in literature \cite{LeakyDNNSideChannel, stealingPredictionAPIs, yuan2020es, deepSniffer}. Therefore, given the vast amount of deployment scenarios it is necessary to alleviate the limitations present within current work, and further study the common associations within extraction attacks across a multitude of deployment scenarios, DL types, and datasets.

PINCH targets the capability of an efficient and automated extraction attack framework able to deploy and evaluate DL model security across heterogeneous DL systems and extraction attack scenarios with design goals of: (1) \textit{Generalizability}, providing a unified platform for the hardware and software used in DL system deployments, and for the attacks that target them. (2) \textit{Configuration \& Automation}, providing a machine independent system to define an attack scenario and deploy it for repeatable experimentation at scale. 

%-------------------------------------------------------------------------------
\section{Threat Model}\label{sec:threatmodel}
%-------------------------------------------------------------------------------
 
The objective of a DL model is to map an input to a provided classification. Given an input, the model propagates through its operators to output a vector of probabilities denoting the confidence of classification associated to the input. The threat models underpinning extraction attacks in this paper are categorized into three aspects: \textit{Model knowledge}, \textit{DL System Environment knowledge} and access to the \textit{Auxiliary dataset}.

\textbf{Model knowledge}. We consider two access types \textit{observed ($M_{o}$)} and \textit{hidden ($M_{h}$)}. With \textit{observed} knowledge, an adversary has access to sufficient\footnote{We deliberately use the term "sufficient" as certain attacks only require a limited sub-set of target model information to succeed.} information of the target model (architecture, parameters) to infer its model characteristics. With \textit{hidden} knowledge, adversary access is limited to API calls to the target model (query, data output from model), with attacks \cite{GanjuInference} assuming that the model architecture is already known to construct a shadow model. \textit{Hidden} knowledge encompasses scenarios whereby a target model is accessed via external API calls commonly found in Machine Learning as a Service (MLaaS). \textit{Observed} knowledge encapsulates information leakage of target model characteristics via attacks such as bus snooping, and side-channel.

\textbf{DL system knowledge}. Knowledge pertaining the DL system environment is used to infer DL model characteristics. Two types of knowledge are considered for the DL environment: \textit{partial ($S_{p}$)}, and \textit{none ($S_{n}$)}. Types denote the adversaries knowledge associated with the target DL environment the target model is executing upon. This includes knowledge regarding DL framework, GPU accelerators, and CPU devices. \textit{Partial} knowledge of the environment enables an adversary to have direct or indirect knowledge about the DL environment, for example, knowing the type of CPU (Intel, AMD), or GPU (Nvidia, AMD). \textit{None} states the adversary has no information regarding the DL environment, and encompasses scenarios whereby an adversary may have no, or not need any knowledge about the DL system.

\textbf{Auxiliary dataset}. Depending on the type of attack, the adversary may require an auxiliary dataset to perform their attack. We consider two scenarios in decreasing order of adversarial "strength": (1) \textit{Partial ($D_{p}$)} where an adversary has some knowledge of target dataset and therefore can obtain parts of the target dataset (e.g. via public knowledge, or staging previous attacks). (2) \textit{No dataset ($D_{n}$)} whereby the adversary has no information regarding the dataset. We assume the attacker has access to open source datasets commonly provided by DL libraries or online repositories \cite{torchvision, CelebA, imagenet}.

\textbf{Overall scenarios}. Considering the model knowledge, DL environment knowledge, and auxiliary dataset a total of 8 distinctive threat models are possible. In the rest of the paper we focus on two: ($M_{h}$, $S_{n}$, $D_{p}$), and ($M_{o}$, $S_{p}$, $D_{n}$). As the scenarios are tailored to extraction attacks 6 scenarios are omitted due to their indifference to extraction success. 

\section{Extraction Attacks}\label{sec:extractionattack}

 \subsection{KnockOffNets (KON)}

KnockOffNets (KON) \cite{knockOffDLs} is an inference attack whereby an adversary undergoes inference upon a target model by querying with a set of images randomly sampled from a \textit{query set} to steal target model parameters and recreate a stolen model. All predictions made by the target model are combined into a new \textit{stolen dataset} containing the previously queried images and stolen prediction confidence values or label pairs. The stolen dataset is then used to reconstruct a new model via a training recreation technique, in which an untrained model of the same architecture as the target is trained on stolen dataset until the desired similarity to the target is reached. The adversary's intention is for the stolen model to be equivalent when compared to the target model within the targets task.

KON leverages the assumptions ($M_{h}$, $S_{n}$, $D_{p}$). The adversary has hidden knowledge access to the target model while being capable of performing inference requests with queries, and does not assume any rate limiting or other inference countermeasures associated with the target model. Inference extraction attacks only use the API to access the DL target model which is abstracted away from the underlying DL system, meaning no DL system knowledge is required. The adversary has partial auxiliary dataset knowledge about the underlying target model architecture and training dataset used to therefore establish a query set to be used during the attack.

\subsection{DeepSniffer (DS)} \label{sec:deepSniffer}

DeepSniffer (DS) focuses on utilizing leaky information from the GPU to infer target model architecture \cite{deepSniffer}. DS captures 4 kernel metrics during operator execution; \textit{execution time ($Exe_{Lat}$), read volume ($R_{V}$), write volume ($W_{V}$), and I/O output volume ($I_{V}$ / $O_{V}$)}, to understand the relationship between operators and variance of metrics. From this relationship, DS can infer one of seven operators within a target model architecture; \textit{Conv, ReLU, BN, Pool, Concat, Add, and FC kernels} via a pre-trained DL model trained upon previous examples leaked by the GPU, called the \textit{DS model}. There are two attack stages: (1) \textit{Attack Staging}: Whereby DS gathers required GPU metrics during target model execution. (2) \textit{Exfiltration:} DS uses the gathered data to undergo architecture prediction via a previously trained DS model.

We make the following assumptions ($M_{o}$, $S_{p}$, $D_{n}$): The adversary observes knowledge about the target model via architectural hints exposed within the GPU that the target model is executing upon. We assume the adversary has partial knowledge of the DL system, and the capability to access low level system functionality including capturing stream of memory and PCI metrics for CPU/Memory $\rightarrow$ GPU communication of the target model to infer kernel metrics through GPU profiling tools such as NVPROF \cite{nvprof}. Knowledge pertaining to the target models ML framework is also considered due to the attack staging requirements. Finally, the adversary must be capable of performing inferences upon the target model. Auxiliary dataset knowledge is not required as activating the networks operators is the focus, not the models prediction, for which data outside the auxiliary set can be used.

\subsection{DeepRecon (DR)} \label{sec:deepRecon}

DeepRecon (DR) \cite{deepRecon} is a side-channel extraction attack that gathers target model architecture information by using information leakage from a device's CPU L3 cache. DeepRecon extracts eight DL operators (\textit{Conv, MatMul, Softmax, Relu, MaxPool, AveragePool, Merge \& Bias}) by associating them with symbols from the target model framework binary and identify their execution by starting a co-located programme to monitor L3 cache. In the case of a CPU attack, Flush+Reload \cite{flushReload} is used to flush the CPUs L3 cache to observe which symbols repopulate the cache on the assumption that frequently occurring symbols belong to a target executing DL process. Dimensional reduction techniques, such as Principle Component Analysis (PCA), can then be used to cluster gathered symbols during a model execution, making it possible for an adversary to infer the architecture of a model by comparing reduced dimensions to that of other known models.

DR leverages the assumptions ($M_{o}$, $S_{p}$, $D_{n}$): The adversary has observed model and partial DL system knowledge whereby it is known that targeted systems are vulnerable to Flush+Reload. Similarly to DeepRecon, auxiliary dataset knowledge is not required. It is assumed that: (1) An adversary is capable of launching co-located user-level processes on the host of the target model. (2) The target and attacking processes use the same DL framework binaries, to associate symbols with DL operators. (3) The adversary knows which CPU architecture is in use, as Flush+Reload is an Intel exploit.

\subsection{MiFace (Inversion Attack)}

To demonstrate the frameworks ability to enable further attacks to be staged upon extracted models, we implemented the \textit{MiFace} model inversion attack by Fredrikson \textit{et al.} \cite{MiFace}. Model inversion is a privacy violating attack whereby an adversary with access to an inference API seeks to reconstruct a representative example from each class within the DL model. The consequence of such an attack is the ability for images of trained classes within a model to be extracted. For each class within the target model, the adversary performs back-propagation over target model parameters to optimize the input sample so that the corresponding class posterior exceeds an established threshold. An input sample can be a randomly generated image, or another initialization technique denoted via an adversaries knowledge of the DL model.

Model inversion leverages the assumptions ($M_{h}$, $S_{n}$, $D_{p}$): It is assumed that the adversary is targeting a model with hidden knowledge, requiring the capability to perform prediction queries on feature vectors targeted by an adversary. As seen previously in \ref{sec:deepSniffer} and \ref{sec:deepRecon}, no DL system environment knowledge is required as model inversion uses the inference API which provides abstraction from the DL systems software and hardware. Furthermore, partial knowledge of model classes is required: with a facial recognition model, the adversary indirectly knows the model responds positively to faces, and the adversary requires access to an auxiliary dataset providing input initialization values.

%-------------------------------------------------------------------------------
\section{Framework Design} \label{sec:frameworkdesign}
\begin{figure}
\begin{center}
\includegraphics[width=\linewidth]{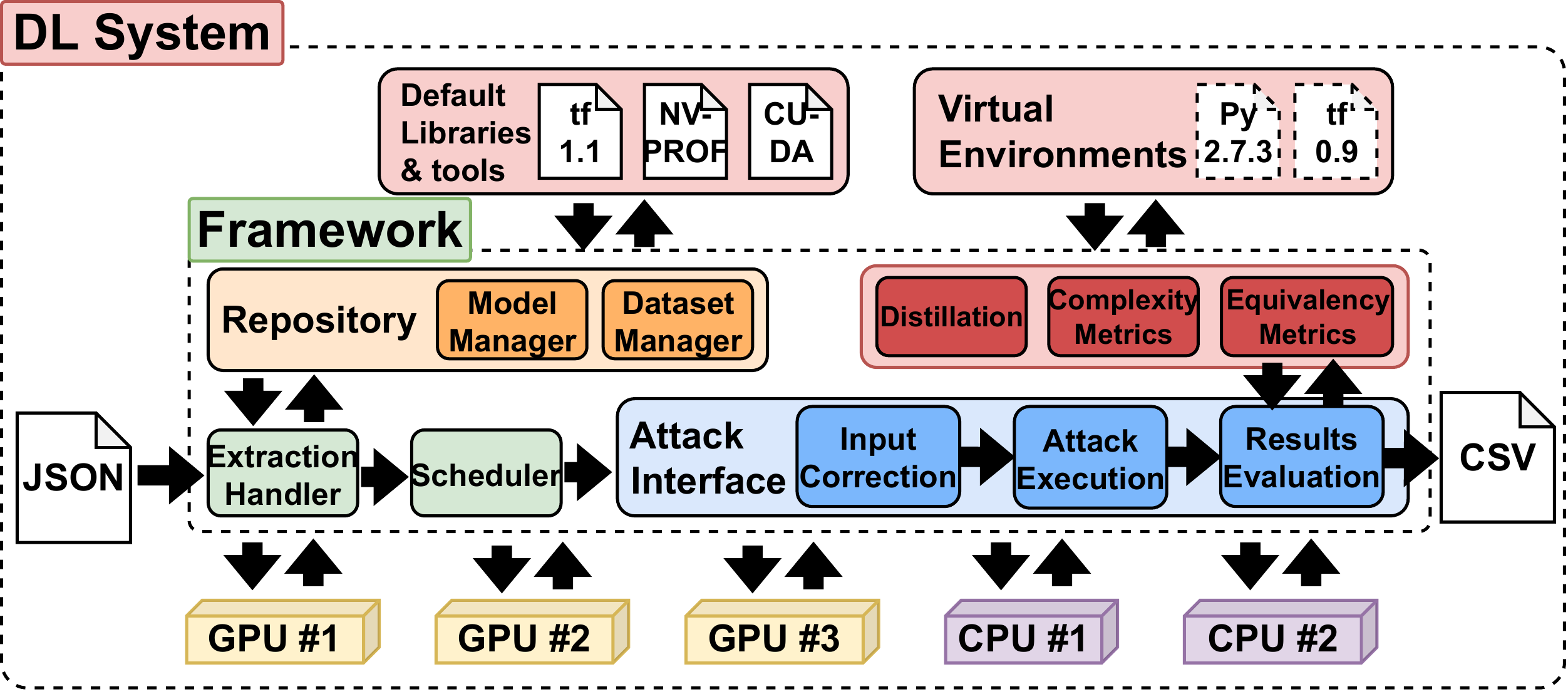}
\end{center}
\caption{\label{fig:components} \textbf{Extraction framework system model and components.}} 
\end{figure}

%-------------------------------------------------------------------------------
The objective of PINCH is to simplify and generalize the process of executing adversarial attacks, facilitating the exploration of associations between attacks, DL model characteristics, and DL systems. The framework accomplishes this by creating interfaces to standardize model inputs, data sets and software environments, providing compatibility for the execution of attacks. Additionally we enable readily reproducible configuration and automation of attack scenarios to gather insights, with straightforward deployment into a DL system without coding or complex build processes. Figure \ref{fig:components} depicts PINCH and its five components: \textit{Scheduler}, \textit{Extraction Handler}, \textit{Attack Interface}, \textit{Model Manager}, \textit{Results \& Metrics}, and \textit{Repositories}.

%-------------------------------------------------------------------------------
\subsection{Components} \label{sec:components}
%-------------------------------------------------------------------------------

%---------------------------
%% %---------------------------

%% %--------------------------- 30/08/2022 -- Re-writing to be more technically dense
\textbf{User Interface}. PINCH was designed for both \textit{Command Line Interfaces (CLI)} and \textit{Browser Interfaces (BI)}, and can readily interface with established AI/ML/DL pipelines. Internally, extraction attack scenarios are stored as JSON objects and are parsed to configure the module pipeline. Single or multi-stage scenarios are passed to the attack function and the results returned within 10 LoC in Python using the CLI. The BI was implemented using a ReactJS front end and Flask web server framework \cite{reactjs, flask}, providing the same facilities but with GUI features, such as drop-down attack scenario configuration and generated results visualizations.

\textbf{Extraction Handler}. We implement the pipeline software design pattern \cite{pipeline} by instantiating components required for an attack scenario and having unidirectional data flow, with the \textit{extraction handler} being the pipeline orchestrator. Given large datasets and models are I/O and memory intensive to load and unload, the extraction handler requests the dataset and model managers to begin loading resources from disk immediately when executed to reduce pipeline latency. Additionally, software environments (libraries, frameworks, interpreters) are preemptively created for a given attack to execute within, either referencing software installed on the DL system, or through the use of virtual environments. Alternative Python library versions are created and accessed using \textit{venv} \cite{venv}, creating lightweight site directories isolated from the default system packages. Attacks with more complex dependencies and build processes, such as DeepRecon, are assigned and run in a containerized environment via docker.
%, with a volume shared on the host machine for data access.

\textbf{Scheduler}. We enable multiple attack scenarios to be executed concurrently on a DL system through a scheduler. A heuristic method assesses the attacks, the currently available resources (\textit{e.g}. GPUs and memory), and decides whether the attacks can be ran in parallel. Once agreed, the DL frameworks are configured to use the assigned resources and attack interfaces components instantiated for each attack. Parallel compute time for KON is $ \frac{1}{n} $ with $n$ GPUs installed in the DL system, compared to linear execution. Side-channel snooping attacks are not run in parallel, as the operations of other executing attacks (\textit{noise}) may provide unsatisfactory results.

\textbf{Attack Interface}. The \textit{attack interface} creates a valid input configuration for executing a given attack scenario. As mentioned in Section \ref{2C}, successful attack execution is highly dependent on input data being syntactically correct. The attack interface implements stub methods that manipulate the attack input from intermediate representations to the standard compatible for the attack depending on scenario. This protects from crash-stop failures caused by fragile input errors regardless of extraction attack and model architecture. The interface wraps the attack execution calls, recording queries and responses or predicted architectures, and performs attack-contextual evaluation \textit{e.g}. calculating model extraction fidelity and similarity methods\cite{CanCorrelation}.

\textbf{Dataset Manager}. Inefficient loading techniques in existing attack frameworks makes testing contemporary datasets often exhaust system memory by attempting to load entire datasets simultaneously. This was resolved by creating a \textit{loader} that automatically splits requested datasets into chunks and loads them progressively into memory on demand, and by limiting the retrieved dataset objects to the number of queries set to execute, rather than the entire set.

\textbf{Model Manager}. The model manager fulfills requests by attacks to load models, while providing for compatibility and reduced user involvement. PINCH maintains a repository consisting of online \cite{torchvision, keras} and local DL models and their checkpoint history. We built a server to serve models by framework, architecture, stage of training and subsets trained classes, additionally implementing automatic transfer learning capabilities \cite{transferLearning} to: (1) train variants of existing model architectures with compatibility for new datasets, and 2) train models used in attacks on targeted subsets of a models classes.

%-------------------------------------------------------------------------------
\section{Experimental Setup} \label{sec:setup}

%-------------------------------------------------------------------------------

%\subsection{Experiment Configuration}

Experiments consisted of studying 21 state-of-the-art DL architectures trained across four benchmark datasets, creating a total of 92 target models. These target models were exposed to three extraction attacks (see Section \ref{sec:extractionattack}) and one model inversion attack. Complying with threat models described in Section \ref{sec:threatmodel}, datasets are separated into an auxiliary and testing sets for fidelity comparisons.

Using PINCH, target models and corresponding dataset were automatically deployed into specific DL framework and hardware devices (Section \ref{sec:hardware}). Next, a configured extraction attack was launched against the target model (Section \ref{sec:targetModels}) and dataset (Section \ref{sec:datasets}). On attack completion, we used our framework to extract and collate results to analyze, research, and study extraction attacks within Sections \ref{sec:evaluation} and \ref{sec:discussion}.

\subsection{Hardware \& Software Setup} \label{sec:hardware}

Experiments were conducted on multiple hardware platforms. KON -- whose effectiveness is reported to be independent of hardware device type -- \cite{knockOffDLs} was evaluated on a Nvidia TESLA V100, and Intel Xeon Gold 5218. Extraction attacks designed to target particular software and hardware device vulnerabilities (DS, DR) were studied across three Nvidia GPUs (TESLA V100, GTX 1080, GTX 970) and three Intel CPUs (i5-3470, i7-4770 and i7-6850k). These devices were selected to study extraction attack effectiveness when exposed to different hardware dimensions of GPU architecture (Maxwell, Pascal, Tesla),  GPU Compute schedule capability, CPU generation (3rd, 4th, 6th) and CPU cache size (6MB, 8MB, 15MB). All experiments used Ubuntu 20.04.2 with CUDA 11.3, and were performed on ML frameworks PyTorch 1.11.0, and TensorFlow 1.10.0 / 2.10.0.

\subsection{Datasets} \label{sec:datasets}

Experiments leveraged four datasets selected based on their complexity (class size, image size, number of channels), and their observed impact upon API-based extraction attack on inference models \cite{mldoctor}. Datasets include: 

%PG - Not sure how the below text commented out relates.
%training splits are denoted by the selected dataset:

%CSG 7.539
\textbf{MNIST}. \cite{mnist} 60,000 training and 10,000 test greyscale images with 28x28 input size and 10 classes. Images contain white hand--written numbers on a black background.

%CSG 79.67
\textbf{CIFAR100}. \cite{cifar100} 50,000 training and 10,000 test colored images with a 32x32 input size and 100 classes. Images represent photos of animals, buildings, and vehicles.

\textbf{CelebA}. \cite{CelebA} 200,000 celebrity face greyscale images with an 218x178 input size, associated with 40 attributes. We selected 10 faces out of 10,177 identities, as  inclusion of this dataset is focused on investigating further attack staging described in Section~\ref{sec::furtherAttacks}.

%CSG 648.98
\textbf{ImageNet}. \cite{imagenet} 14 million colored images with an 224x224 input size and 1000 classes. In our experiments, we used a subset of ImageNet providing 80,000 training and 20,000 test images (which we simply refer to as ImageNet).

Two types of datasets are used to undergo and evaluate extraction attack effectiveness. (1) \textit{Query dataset:} A collection of inputs that an adversary can use to extract stolen classified labels from a target model. The query dataset size represents the amount of queries an adversary can perform. (2) \textit{Testing dataset:} Used to evaluate extraction fidelity of the stolen model compared to the target model. Test sets are derived from the selected dataset test images, and thus are related to the target model trained dataset. 

\subsection{Target Models} \label{sec:targetModels}

Our study utilizes 21 DL model architectures~\footnote{Appendix Table \ref{appendix:targetArchitectures} details the full model types and configurations.} across three data sets described in Section \ref{sec:datasets}, with the exception of CelebA dedicated to exploring the impact of further attack staging. Architectures selected for evaluation were chosen based on diversity of parameter size, model family, commonality within extraction literature, and also includes more modern models such as ConvNext, RegNet, and ViTB16 \cite{convnextArchs, regnet, vitB}.

%MobileNetV2 evaluate the trade-offs between models through number of operations measured by multiply-adds (MAdd) \cite{MAdd}. We additionally, to model parameters, utilise this metric to measure the complexity of the architectures. \hl{\textbf{W - Adding MAdd here.}}

Target models were acquired via online repositories or trained locally. Online target models were sourced from TorchVision for KON and DS \cite{torchvision}, or Keras for DR \cite{keras}, which provide pre-trained ImageNet weights upon a given architecture. MNIST and CIFAR target models were trained via a transfer learning approach where pre-trained ImageNet models were re-trained to learn the new dataset. Training was performed with mini-batch size set to 10, and a cross-entropy loss function with a learning rate of 0.01 using a v100 GPU \cite{v100GPU}. Target models were trained for 3 and 40 epochs for MNIST and CIFAR, respectively following similar training models in literature \cite{mnist, cifar100}.

\subsection{Extraction Attacks} \label{sec:extractionAttacksSetup}

\textbf{KnockOffNets (KON)}. We used MNIST, CIFAR100, and ImageNet as query datasets for target models. Stolen model training leveraged identical training setup in Section \ref{sec:targetModels} with epochs set to 10, 20, and 100 for MNIST, CIFAR, and ImageNet, respectively and were selected due to differences in dataset complexity and success indifference \cite{mldoctor}. The number of maximum queries for each attack relates to the training set size or is chosen due to overfitting with indifferent success, therefore the only reduced dataset was MNIST which used 10,000 queries, while CIFAR, and ImageNet both used their maximum training size. For the purposes of generalization, all query dataset input image sizes were transformed to 224x224. We also investigated the impact of dataset class sizes by randomly selecting classes available from the dataset to create smaller subsets which were trained across all architectures.

\textbf{DeepSniffer (DS)}. Kernel metrics were collected using NVPROF to profile a model during a single inference similarly to Hu \textit{et al.} \cite{deepSniffer}. Each target model was extracted 25 times to measure variation in extraction success across runs and then additionally repeated across all evaluated GPUs (GTX 970, GTX 1080, Tesla V100). This collectively totals 750 results for running DS across the system.

\textbf{DeepRecon (DR)}. Target models were deployed within two containerized software environments. The first, identical to \cite{deepRecon}, was configured to perform 10 inferences each on 10 models, accounting for stochastic interference from the operating system. When inference begins, symbol extraction starts, and the detected symbols collected. Each model performed 10 inferences 100 times, totaling 10,000 results across per machine. Thresholds of 200 (default) were reported for all runs, as alternative thresholds (mean value using the Mastik FR-threshold function) were found to have no impact upon results. Outlined in Section \ref{sec:deepRecon}, PCA was used to fingerprint the models, by using computed components to train and evaluate a KNN-classifier to classify the model architecture, and family from a given symbol result set. A model's depth and architecture could be predicted (an \textit{exact classification}), but the predicted model family was also recorded (\textit{family classification}). The second environment ran a version of DeepRecon compatible with TensorFlow 2.10, allowing more models to be tested. Certain cache symbols, specifically MatMul, were found to not be compatible with TF 2.10, though were not replaced to avoid modifying the attack.
%with an 85\% and 15\% split 
 
\subsection{Evaluation metrics} \label{sec:evaluationMetrics}

 \textbf{Attack success}. Extraction \textit{fidelity} is a metric is widely used within extraction literature \cite{stealingPredictionAPIs, knockOffDLs} to measure attack success, whereby the characteristics of the stolen and original model are directly compared by using the Top1-accuracy of predicted classified labels (KON) or architecture prediction (DS). Additional metrics of relevance collected including number of queries (KON) and model accuracy.
 
 \textbf{Complexity Measurement}. To study the relationship between adversarial attack and target model complexity, we utilize two different techniques to quantify dataset and model architecture complexity. Dataset complexity was measured using \textit{Cumulative Spectral Gradient (CSG)} \cite{CSG} that projects extracted dataset features into lower dimensions and calculates class overlap using a Monte-Carlo method, capturing dataset separability, with higher values denoting higher complexity. Architecture complexity was measured by multiply-add (MAdd) \cite{MAdd}, a technique whereby the total number of multiplication operations is computed via operator feature maps sizes within the DL model.   

\section{Extraction Attack Evaluation} \label{sec:evaluation}

%\hl{ST- Me and Will have discussed a feature of the paper we need to address in the narrative: how do we explain that we have made it so all attacks \textbf{WORK}, but are not necessarily \textbf{SUCCESSFUL}. For example, we can test KON or DS with pytorch or tensorflow, but for DS tensorflow gives poor results. It's not our job to fix the attack to be \textbf{successful} on the latest models, but to make sure it can \textbf{work} in as many scenarios as possible.}

\begin{figure}[t!]
\begin{center}
\includegraphics[width=\linewidth]{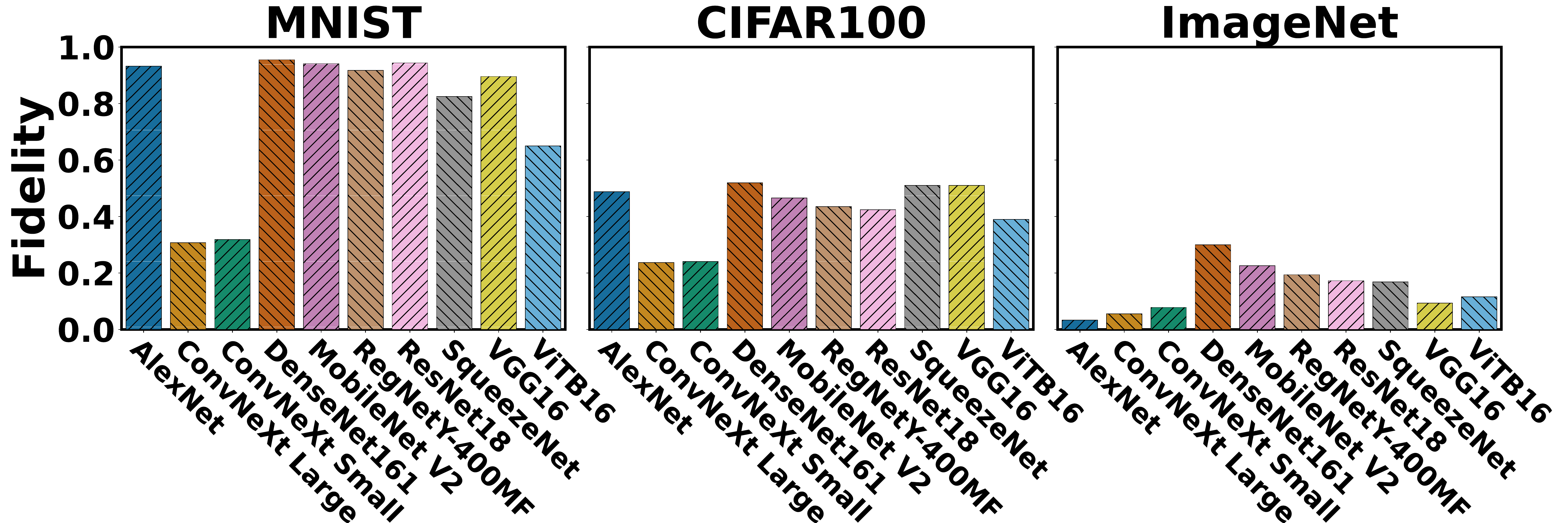}
\end{center}
\caption{\label{fig:extraction} \textbf{KnockOffNets extraction results.} Extraction success across architectures and datasets. CSG: MNIST (7.54), CIFAR100 (79.67), ImageNet (648.99).}
\end{figure}
%% %---------------------------

%---------------------------
\begin{figure}[t]
  \begin{subfigure}{0.23\textwidth}
    \includegraphics[width=\linewidth]{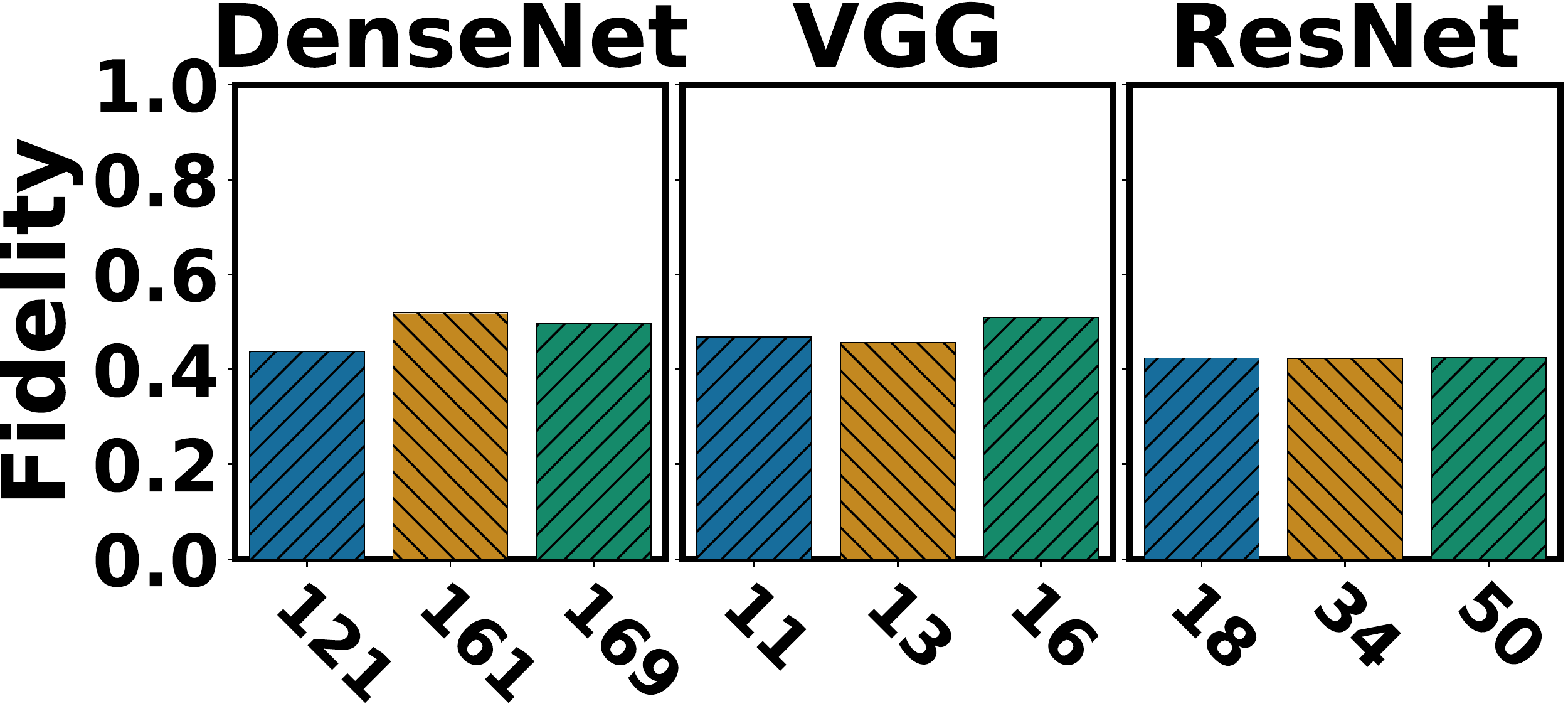}
    \caption{KnockOffNets} \label{fig:depth_knockoff}
  \end{subfigure}%
  \hspace{0.02pt}
  \begin{subfigure}{0.23\textwidth}
    \includegraphics[width=\linewidth]{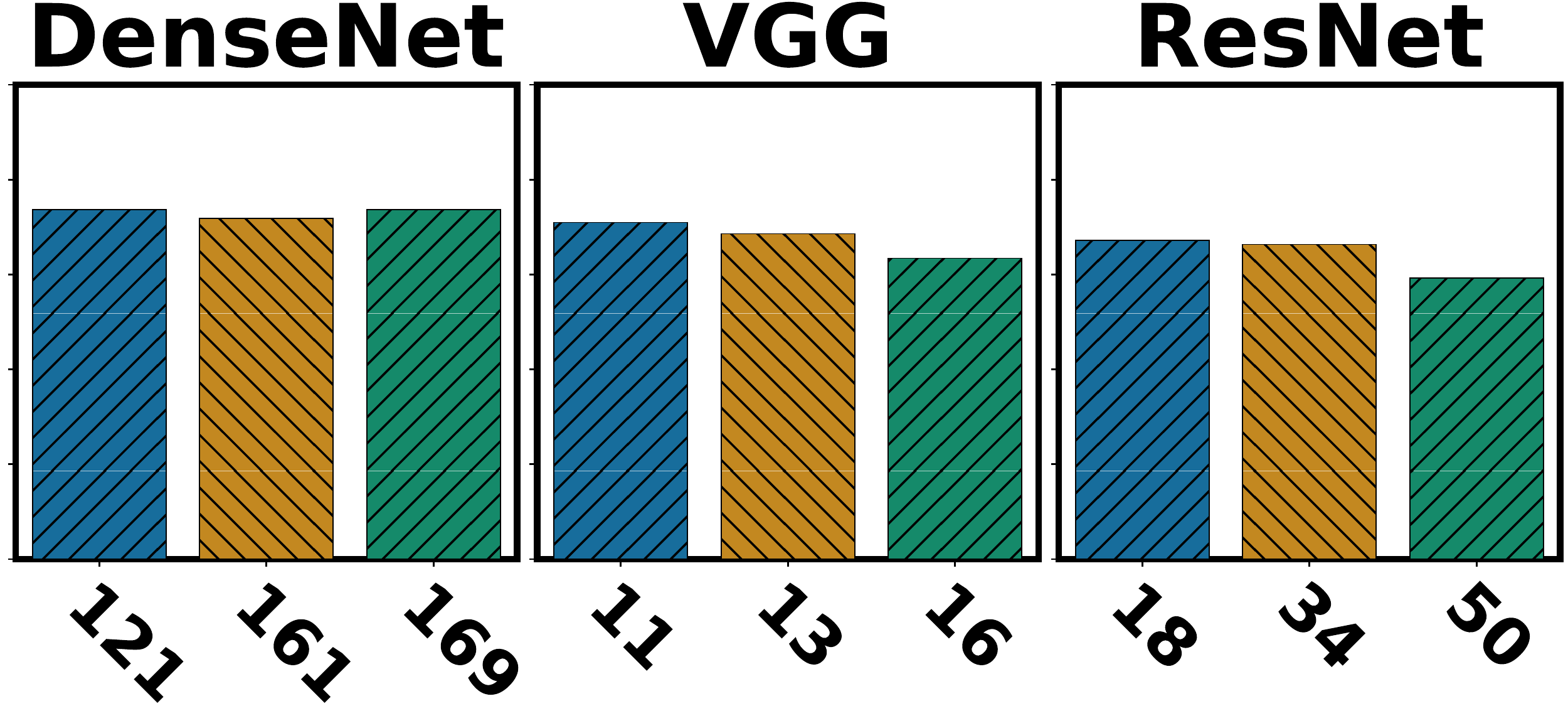}
    \caption{DeepSniffer} \label{fig:depth_deepsniffer}
  \end{subfigure}%

\caption{\label{fig:depth} \textbf{Architecture family depth comparison.} Fidelity variance with models of different depths/parameters using same model architecture family.}
\end{figure}
%---------------------------

\subsection{KnockOffNets (KON)}

KON exhibited varying success across different target model and dataset combinations as shown in Figure \ref{fig:extraction}, with DenseNet161 achieving highest overall success in MNIST (0.95), CIFAR100 (0.52) and ImageNet (0.29). We identified multiple influences on attack success from model architecture, dataset complexity, class size, and number of queries.

\textbf{Model architecture}. We observed that target models (AlexNet, DenseNet161, VGG16) leveraging more conventional CNN architecture achieved higher success. In contrast, state-of-the-art target model architectures (ConvNeXt Small, ViTB16, RegNetY-400MF) reported lower success, with the transformer model ViTB16 exhibiting the lowest fidelity (0.0 -- 0.18). As shown in Figure \ref{fig:depth_knockoff}, we determined that the number of architecture layers for target model family had minor impact, with minimal variation in extraction fidelity across families. Such architectures exhibit exceedingly complex high amount of trainable parameters and MAdd (Table \ref{appendix:deepReconTestSuite}). Therefore, successfully extracting target models via attacks requiring exact architecture knowledge (such as KON) are increasingly difficult for models with large training requirements (queries and time). Such findings indicate that using complex architectures can intrinsically hinder adversary success without considerable effort, similarly to security methods that attempt to exceed adversary effort as a deterrent \cite{mlSecuritySurvey}.

%---------------------------
\begin{figure}
  \begin{subfigure}{0.48\textwidth}
    \includegraphics[width=\linewidth]{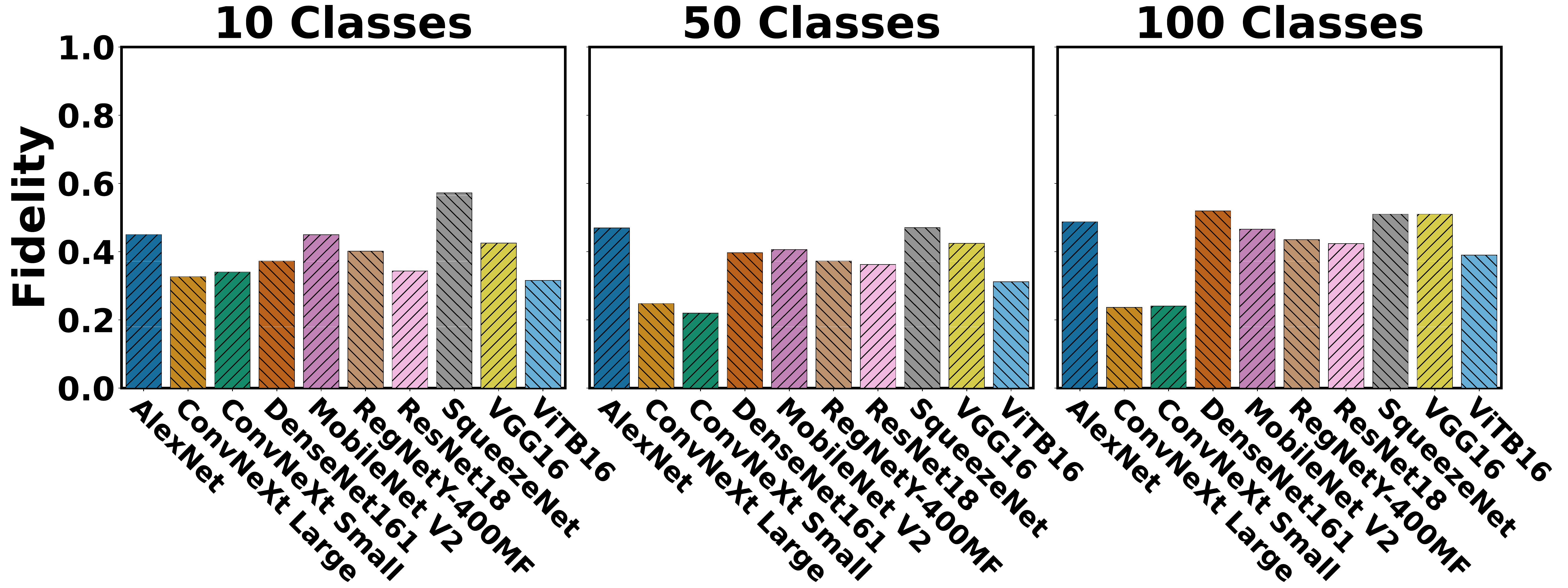}
    \caption{CIFAR100, 50,000 Queries} \label{fig:knockoffcifar}
  \end{subfigure}%
  \\
  \begin{subfigure}{0.48\textwidth}
    \includegraphics[width=\linewidth]{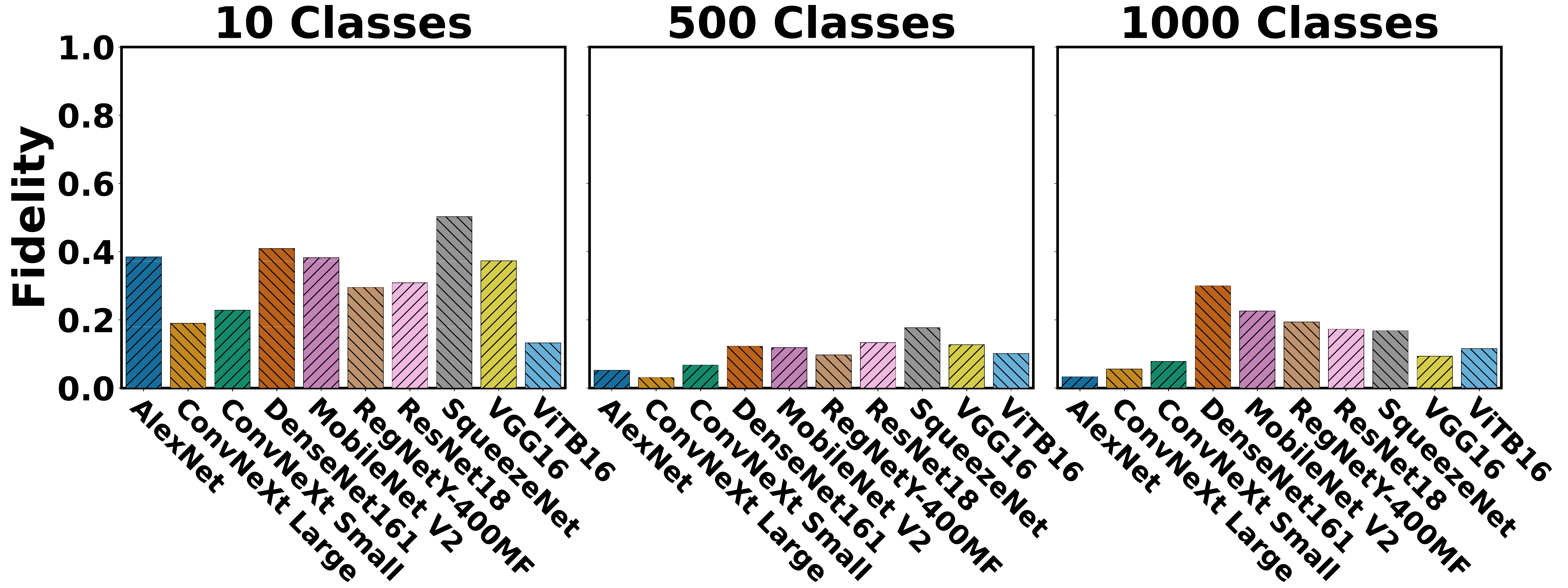}
    \caption{ImageNet, 80,000 Queries} \label{fig:knockoffimagenet}
  \end{subfigure}%
\caption{\label{fig:classSizes} \textbf{Varying class sizes for KnockOffNets attack.} Datasets were reduced to a given class size and were trained via transfer learning.}
\end{figure}
\begin{figure}[t]
  \begin{subfigure}{0.48\textwidth}
    \includegraphics[width=\linewidth]{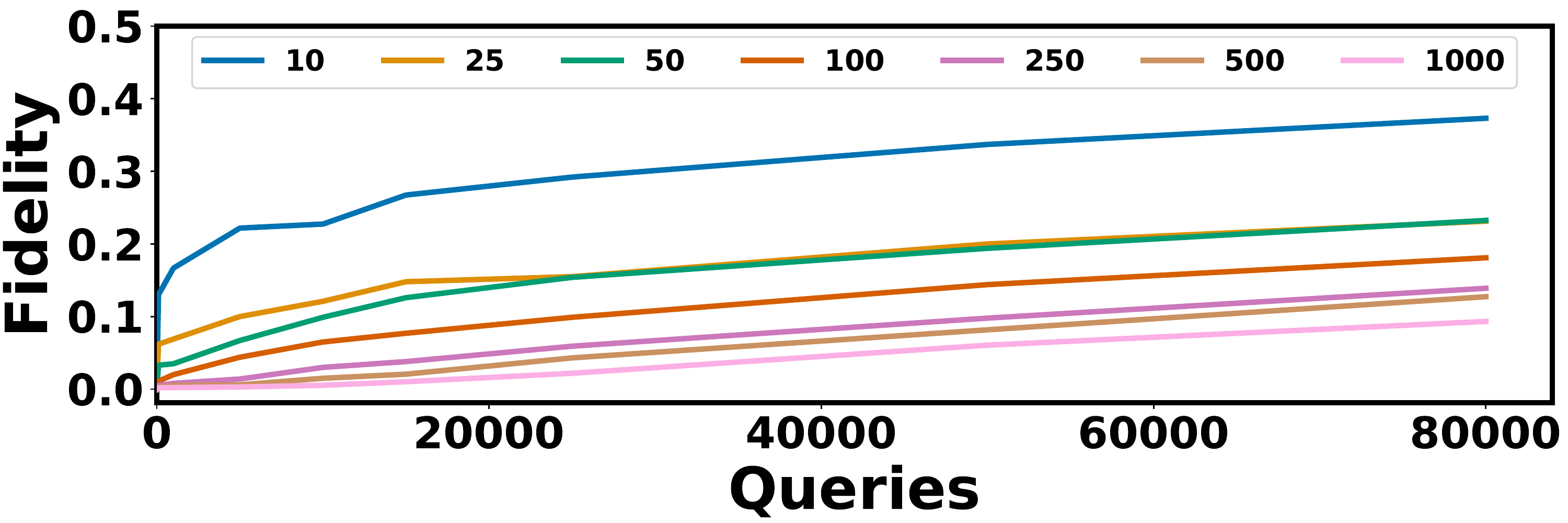}
    \caption{Class size (VGG16, ImageNet)} \label{fig:queries_by_classes}
  \end{subfigure}%
  \\
  \captionsetup[subfigure]{justification=centering}
    \centering
  \begin{subfigure}{0.24\textwidth}
    \includegraphics[width=\linewidth]{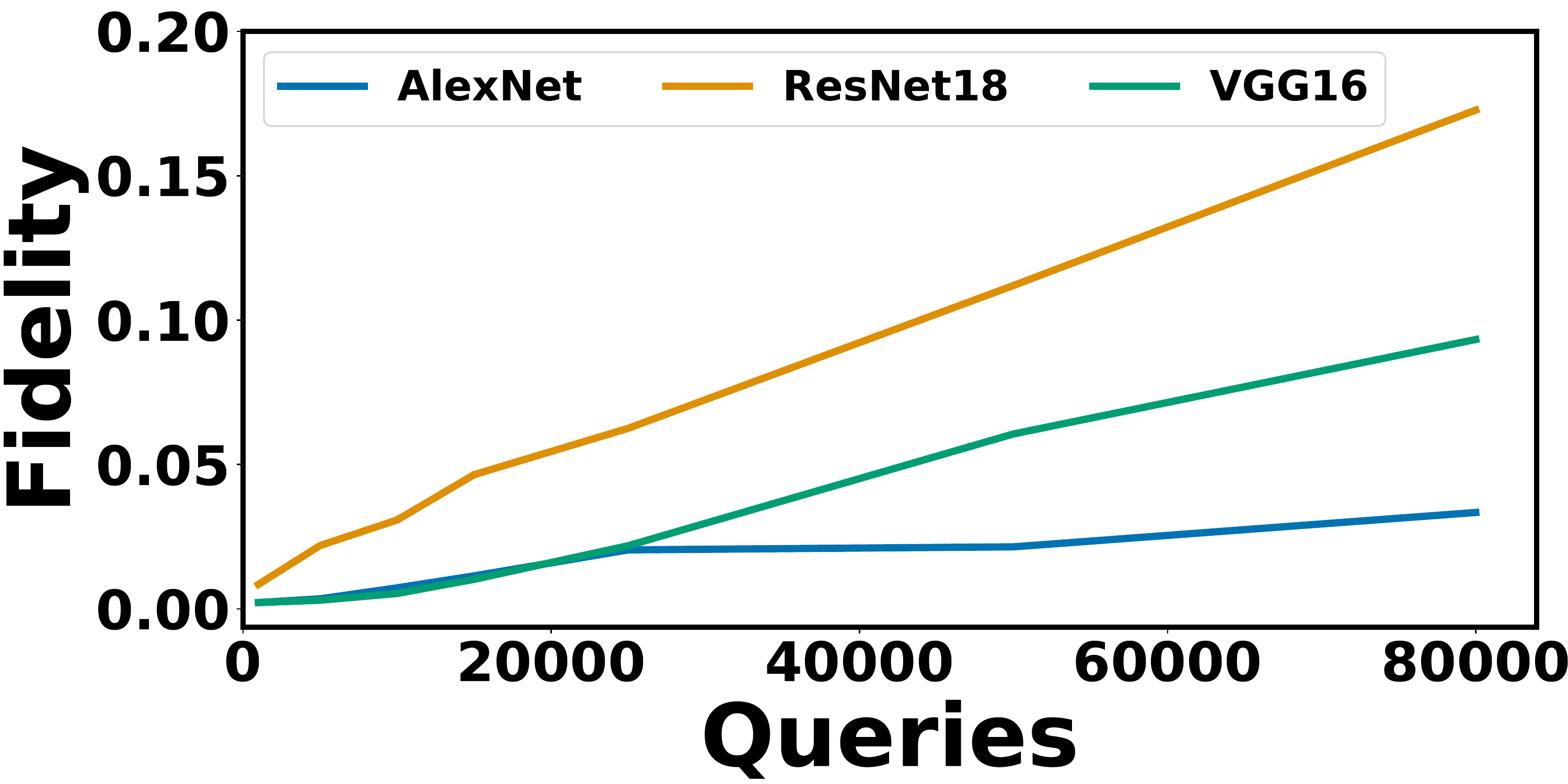}
    \caption{Models\\(1000 classes)} \label{fig:queries_by_model}
  \end{subfigure}%
    \begin{subfigure}{0.24\textwidth}
    \includegraphics[width=\linewidth]{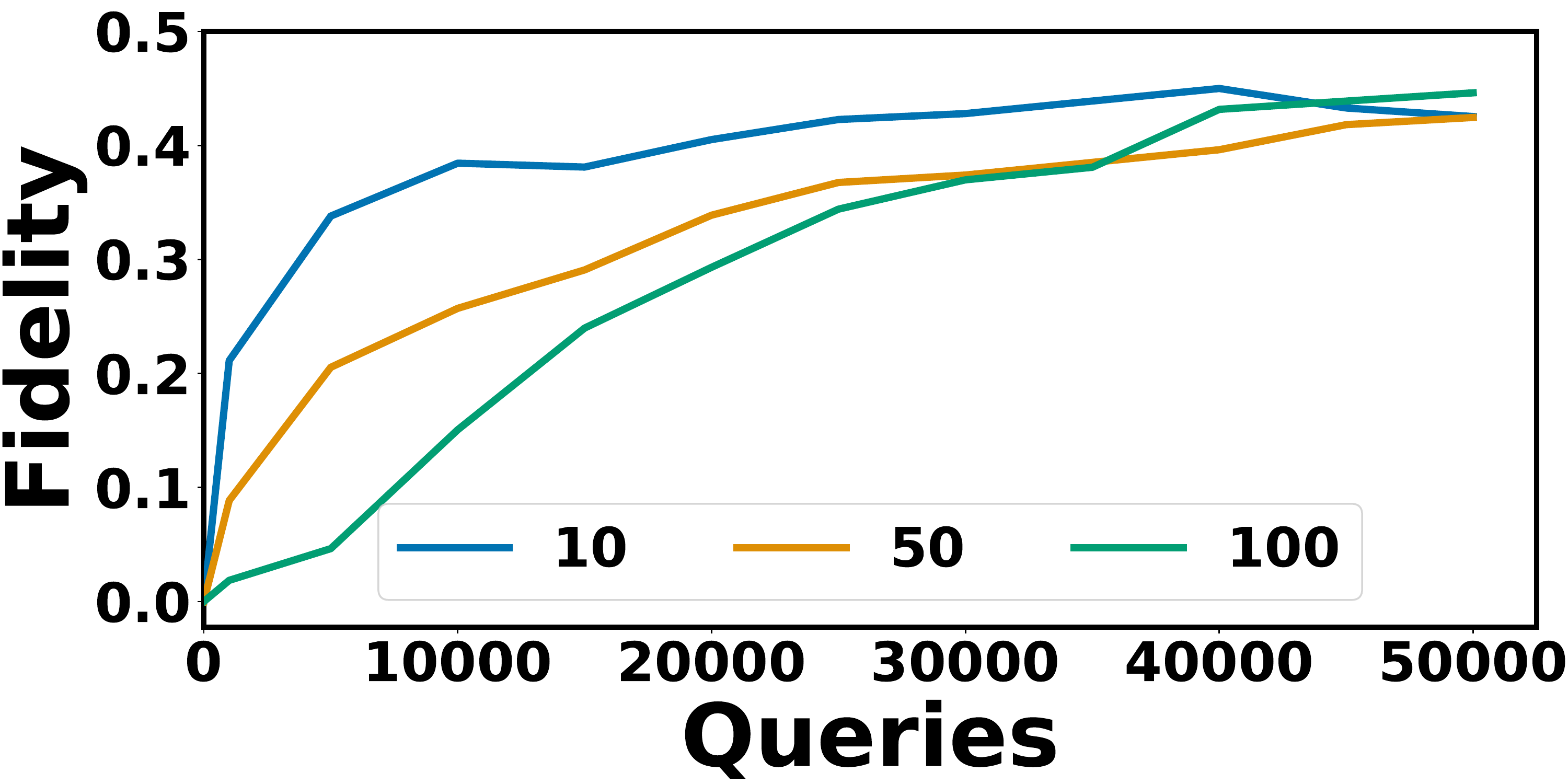}
    \caption{Class size (VGG16, CIFAR100)} \label{fig:queries_by_classes_cifar}
\end{subfigure}%

\caption{\label{fig:query} \textbf{Varying query amounts.} KnockOffNets extraction upon ImageNet and Cifar100 across DL model architectures, various class sizes, and query amounts.}
\end{figure}

\textbf{Dataset complexity}. Target models using MNIST exhibited the highest attack success, with multiple architectures reporting over 0.9 fidelity, whereas ImageNet indicated the lowest attack success between 0.03 -- 0.29 (Figure \ref{fig:extraction}). The reason for such results is due to dataset complexity~\cite{mldoctor}, whereby MNIST is least complex dataset (CSG value of 7.54) in comparison to ImageNet (CSG value 648.99). Increased dataset complexity results in more difficult class generalization and higher likelihood of overfitting \cite{mldoctor}. Figure~\ref{fig:classSizes} demonstrates the impact of dataset class size on attack effectiveness. For instance, increasing ImageNet class size with DenseNet161 from 10 to 1000 resulted in reduced attack success from 0.41 to 0.29. Figure~\ref{fig:queries_by_classes} and \ref{fig:queries_by_classes_cifar} depict that smaller class sizes produce higher fidelity with less queries, and additionally show that fidelity of CIFAR100 is fixed when a model is converged. Intuitively, smaller class size results in reduced dataset complexity, thus allowing a model to generalize faster with less data to achieve higher extraction success.

%However, \hl{CIFAR100 only showed increase to extraction success with ConvNeXt family of models}.

%but show that it doesn't have an effect once training convergences, as seen in CIFAR100 with more than 40,000 queries. Fundamentally, smaller class size results in reduced dataset complexity, thus allowing a model to generalise faster with less data and achieve higher extraction success.

%The difference in success across datasets is due to \hl{training convergence} whereby the number of queries above 40,000 caused convergence across all class sizes in CIFAR100. Evaluating fidelity across varying query amounts we can observe that lower class sizes produce higher fidelity earlier with less queries in CIFAR100 (Figure \ref{fig:queries_by_classes_cifar}), and additionally in Imagenet (Figure \ref{fig:queries_by_classes}).

\textbf{Query number}. As provided in Figure \ref{fig:extraction}, we found across all target models, the number of queries launched [10,000, 50,000, 80,000] for MNIST, CIFAR100, and ImageNet, respectively exhibited the highest impact on attack success whereby datasets of less complexity are stolen quicker. Given the training recreation technique used within KON, whereby more training data would lead (in moderation) to higher success due to greater learning generalization and diversity of data as shown in Figure \ref{fig:queries_by_model}, also reported in~\cite{mldoctor}. Architectures and datasets of higher complexities, such as ConvNeXt upon ImageNet, require larger amount of queries for adequate extraction success. %We specifically highlight that architecture complexity is a stronger factor in this context compared to dataset complexity. As shown in Figure \ref{fig:extraction}, a less complex MNIST dataset providing higher extraction across architectures.

%---------------------------
%% %---------------------------

\subsection{DeepSniffer (DS)}

DS also exhibited varying attack success across all evaluated architectures and DL environments. Across GPU devices, Densenet161 achieved the highest success for GTX 970 (0.71) and GTX 1080 (0.78), and ResNet18 within Tesla V100 (0.71). We determined that extraction attack success was influenced by three main factors; model architecture, ML framework, and GPU environment.

\begin{figure}
\begin{center}
\includegraphics[width=\linewidth]{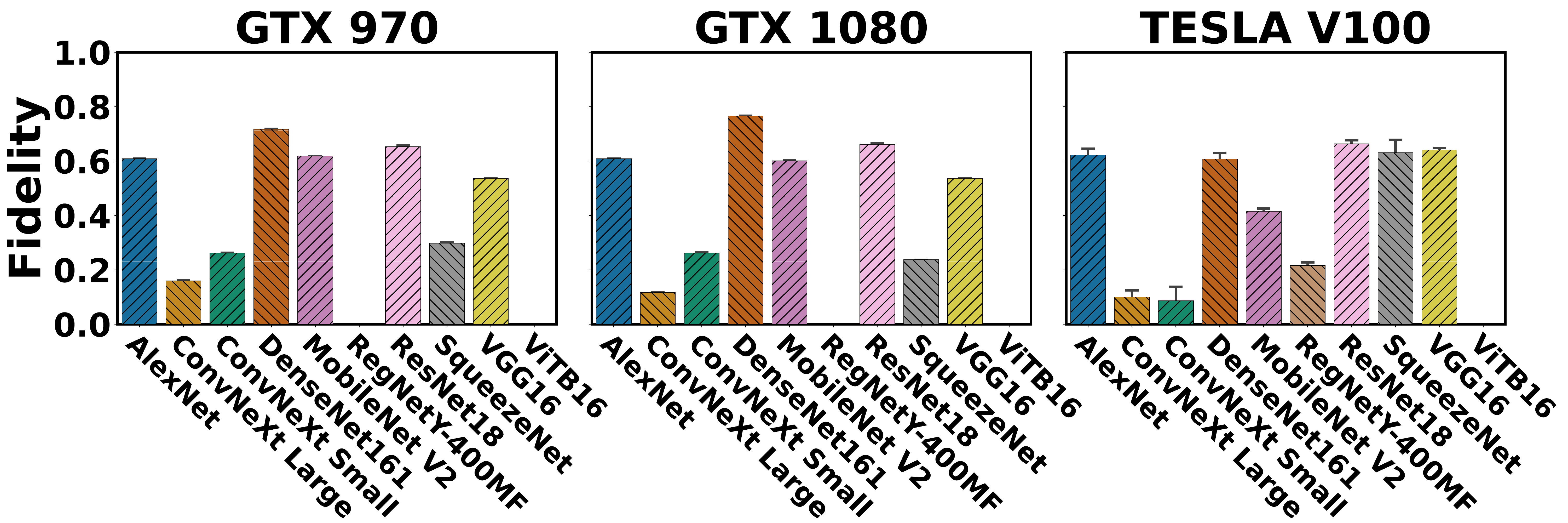}
\caption{\label{fig:deepSniffer} \textbf{DeepSniffer extraction results across GPU devices and various ImageNet DL architectures.}}
\end{center}
\end{figure}

\textbf{Model architecture}. As shown in Figure \ref{fig:deepSniffer}, we observed that DS achieved consistently achieved high attack success across evaluated model architectures. We discovered that DS was ineffective when applied to newer models (ConvNeXt, RegNet, ViTB16) due to including operators and network designs not found within conventional evaluated models (ResNet, VGG, etc) \cite{convnextArchs, regnet, vitB}. ConvNeXt and ViTB16 both implement Gausesian Error Linear Units (GELU) as replacements to widely used Rectified Linear Units (ReLU) \cite{GELU}. ViTB16 additionally uses Transformer specific operators \cite{transformers}, and RegNet introduces a completely new network design paradigm \cite{regnet}. As such, these new operators and architectural approaches cannot be transformed into dimensions recognized within DS (see Section \ref{sec:deepSniffer}) which only capture standard operators within CNNs \cite{deepSniffer}, limiting attack effectiveness for more sophisticated architectures. Moreover, Figure \ref{fig:depth_deepsniffer} demonstrates that deeper models (models with more layers) within the same family resulted in reduced fidelity for VGG and Resnet families, since models of increasing depth include more operators and increased likelihood of layer mis-classification.

\textbf{ML framework}. We found DS was ineffective for DL models using TensorFlow irrespective of target model architecture. From analyzing gathered profiled data across Pytorch and TensorFlow frameworks, TensorFlow generated additional kernel calls not seen within PyTorch. Thus, the increased size of profiled data results in the trained DS classifier predicting an architecture length greater than expected. The existence of such noise can be attributed to low-level framework-level optimizations when kernels execute in comparison to PyTorch. This finding indicates that DS is framework specific, and thus requires training on different frameworks to generalize extraction, and further threat model refinement to enable adversary knowledge of the system framework (Table \ref{appendix:pytorchVTensorflow}).

%---------------------------
%% %---------------------------

\textbf{GPU architecture}. As shown in Figure \ref{fig:deepSniffer}, we observed substantive variance in attack success across architectures for different GPUs, with GTX 970, and GTX 1080 exhibiting similar trends with slight variation. Across all GPUs, we observed clear DS attack ineffectiveness when targeting newer state-of-the-art models (ConvNeXt, RegNet, ViTB16). Of particular interest, RegNetY-400MF, ConvNeXt Small and ConvNeXt Large exhibited low attack success across devices (0.12 - 0.20) and failing to extract for RegNetY-400MF. Such phenomena indicates that architecture attack susceptibility to DS is strongly affected by the GPU device. These architectural differences affect the information leakage gathered by NVPROF, with optimizations and floating-point precision varying between GPUs causing recorded kernel metrics (execution duration, read, write amounts) to change.

%---------------------------
\begin{figure}[t]
  \includegraphics[width=\linewidth]{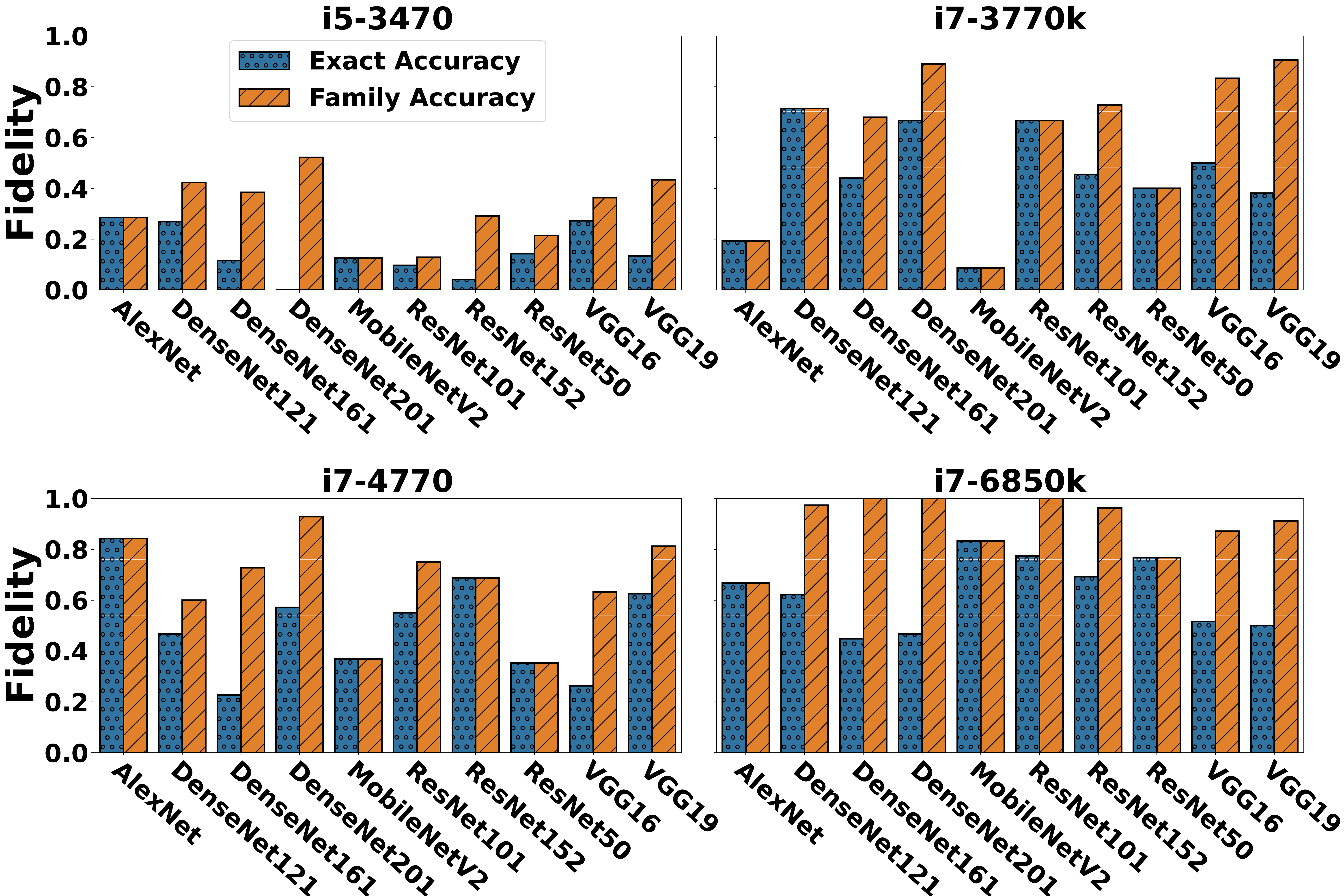}
\caption{\label{fig:deepReconResultsFidelity} \textbf{DeepRecon model architecture \& family prediction.} Classifying model family was more successful in comparison to target model classification, due to similar clustering properties.}
\end{figure}
%---------------------------

%---------------------------
\begin{figure}[t]
  \begin{subfigure}{0.48\textwidth}
    \includegraphics[width=\linewidth]{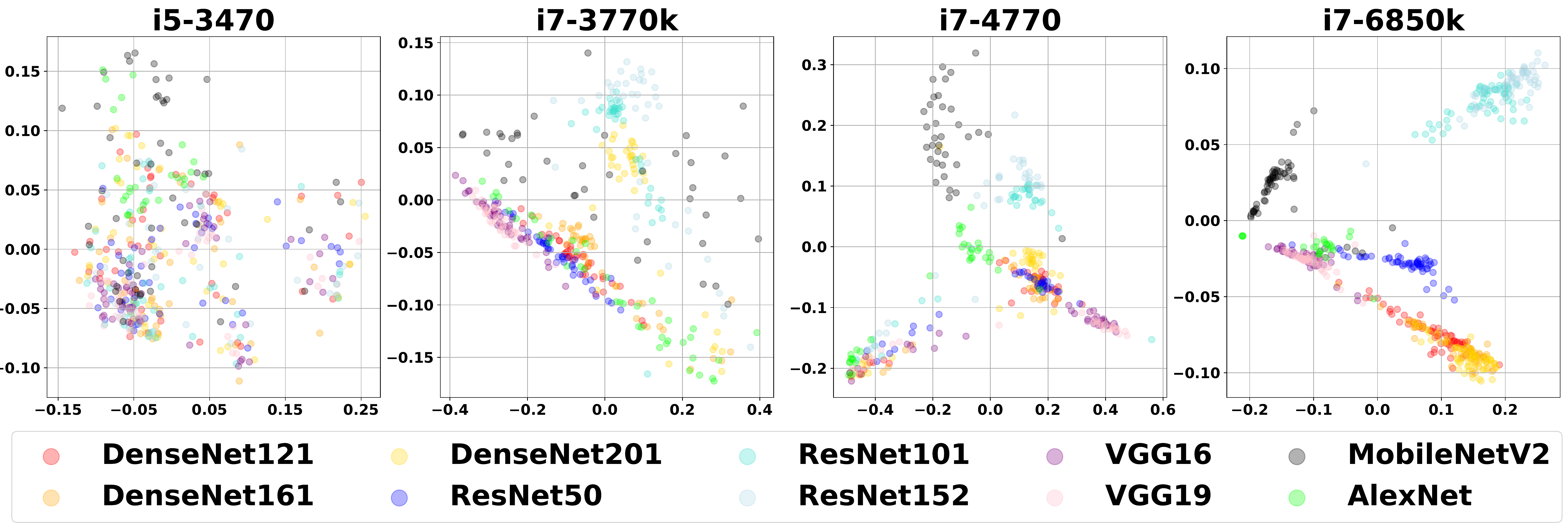}
    \caption{TensorFlow 1.10} \label{fig:deeprecon_tf_1.10_scatter}
  \end{subfigure}%
  \\
  \hspace{0.02pt}
  \begin{subfigure}{0.48\textwidth}
    \includegraphics[width=\linewidth]{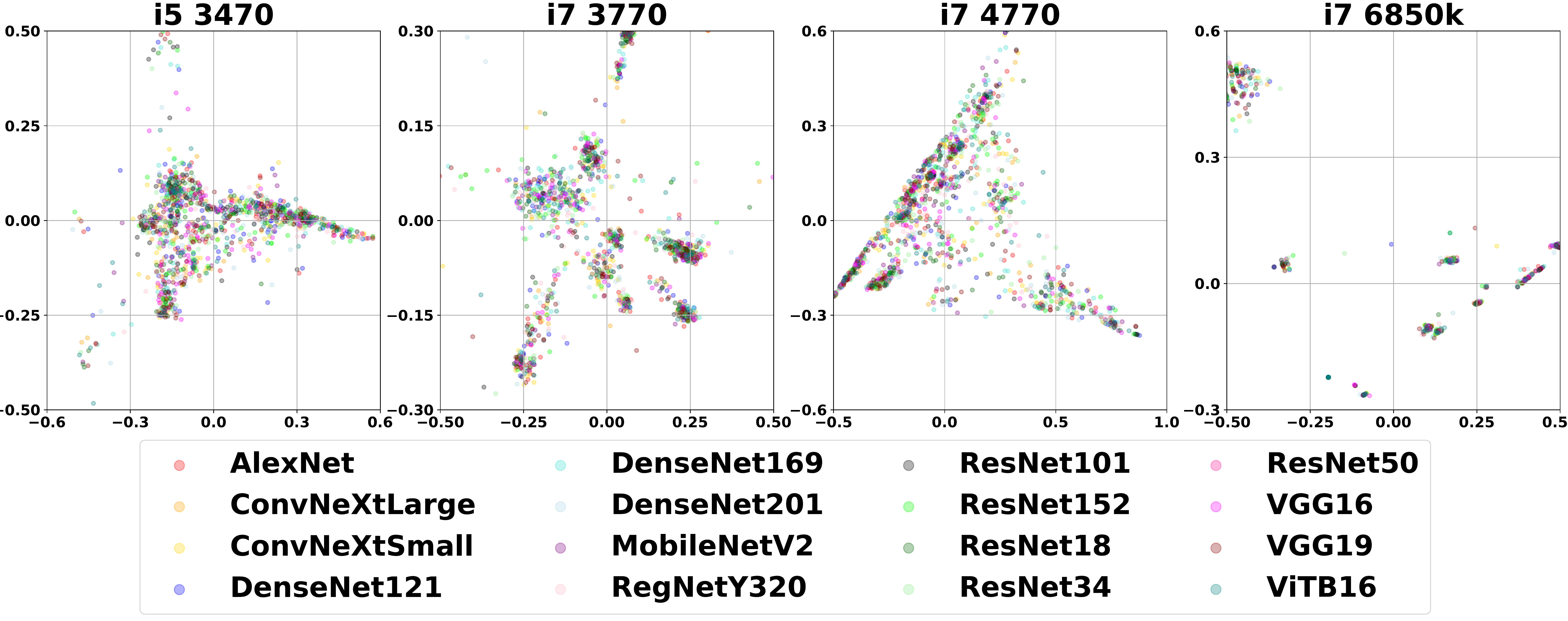}
    \caption{TensorFlow 2.10} \label{fig:deeprecon_tf_2.10_scatter}
  \end{subfigure}%

\caption{\label{fig:deepReconResults} \textbf{Principle component analysis (PCA) for 26,000 inferences using DeepRecon across 16 models, 4 CPUs and 2 TensorFlow releases.} Clustering denotes models with similar proportions of cached operators, clearer clustering indicating similar caching behavior. PCA was used to reduce 8 detected symbols for each inference into 2 dimensions.}
\end{figure}

\subsection{DeepRecon (DR)}

DR demonstrated varying success across hardware platforms evaluated as shown in Figure \ref{fig:deepReconResultsFidelity}. Primarily exploring TensorFlow 1.10, DenseNet (0.835) and VGG (0.766) architectures exhibited high family prediction success on all hardware tested (Figure \ref{fig:deepReconResultsFidelity}), with other families showing a preference for a given CPU (AlexNet: i7-4770, MobileNet: 6850k). We noticed less distinct patterns of success across model architectures of varying depths, e.g. DenseNet depth 161 broadly being classified less accurately than 121 and 201 (Figure \ref{fig:deepReconResultsFidelity}).

\textbf{CPU architecture}. We observed that higher performance CPUs were more vulnerable to DR, with i7-6850k achieving the highest average family classification accuracy (0.836), and i5-3470 lowest (0.316). We attribute this to cache policies and older CPUs being less responsive to timings of Flush+Reload-- e.g. i5-3470 generating large but disparate symbol logs leading to a disturbed PCA result (Figure \ref{fig:deeprecon_tf_1.10_scatter}). Higher performance CPUs (i7-3770k, i7-4770, i7-6850k) reported a higher effectiveness in fingerprinting and classification across model families compared to i5-3470. This is further reinforced in Figure \ref{fig:deepReconResultsFidelity}, with the i5-3470, i7-4770, and i7-6850k averaging 0.316, 0.495, 0.661, and 0.836 across all model families respectively. Findings show cache size does not influence results, as Hong \textit{et al.} \cite{deepRecon} evaluated with a Gen4. 4MB CPU successfully, smaller than our i5-3470 (Gen3, 6MB).

%This is reinforced by classifier results from Figure \ref{fig:deepReconResultsFidelity}, with the i5-3470, i7-4770, and i7-6850k averaging 0.316, 0.495, 0.661, and 0.836 across all model families respectively. Findings also show cache size does not influence results, as Hong \textit{et al.} \cite{deepRecon} evaluated with a Gen4. 4MB CPU successfully, smaller than i5-3470 (Gen3, 6MB).

% i7-6850k (0.836) reported significant improvements over i7-4770 (0.661), which we attribute to it being the most modern CPU (Gen6.) evaluated, with more optimized caching policies that interfere with Flush+Reloads symbol detection less than i7-4770 (Gen 4.). This is strengthened by PCA analysis cumulative explained variance ratio of 92.20\% for i7-4770 and 95.56\% for i7-6850k, (Appendix \ref{appendix:deepReconTestSuite}). Of note is AlexNet and MobileNetV2  models performed poorly on exact and family accuracy on both i5-3470 and i7-3770k, but well on. This finding indicates that models with highly homogeneous operators (AlexNet being primarily convs) can still be classified accurately with an imperfect Flush+Reload result as the proportions of logged operators are not as prevalent.

\textbf{Model architecture \& family}. As shown in Figure \ref{fig:deepReconResults}, we observed target models within model families with distinct operator traits (ResNet; residual operators, DenseNet; dense blocks) successfully fingerprinted, but overlapped with other family members. This is shown further when models were often misclassified individually but with high success within a family. Across all machines, DenseNet exact classification never exceeded 0.518 for any depth (121: 0.518, 161: 0.307, 201: 0.426), but by family averaged 0.835, similarly seen in VGG (Table \ref{tab:deeprecon_averages}). This indicates that families with homogeneous operators built strong fingerprints, but using this to train classifiers for specific depths is insufficient.

\textbf{TensorFlow framework} TensorFlow 2.10 displayed a general decrease in exact and family classification accuracy across i5-3470, i7-4770 \& i7-6850k systems and models, which we attribute to the missing MatMul operator decreasing the reducible dimensions for component analysis. Conversely, i7-3770k performed unexpectedly well compared to other systems, and demonstrated high accuracy on ConvNeXt Small/Large, RegNetY320, and ViTB16 (Appendix \ref{fig:deeprecon_accuracies_tf2}).

% The Inception Family (XCeption; DSC operator, Inception ResNet; residual operators, Inception V3) feature operators not included in other members of the family and hence built weaker fingerprints, intuitively showing lower average family success (0.65) on i7-4770 \& i7-6850k than DenseNet (0.78), ResNet (0.81) and VGG (0.76). 

\subsection{Adversarial Attack Staging} \label{sec::furtherAttacks}

%---------------------------
\begin{figure}[t]
  \begin{subfigure}{0.23\textwidth}
    \includegraphics[width=\linewidth]{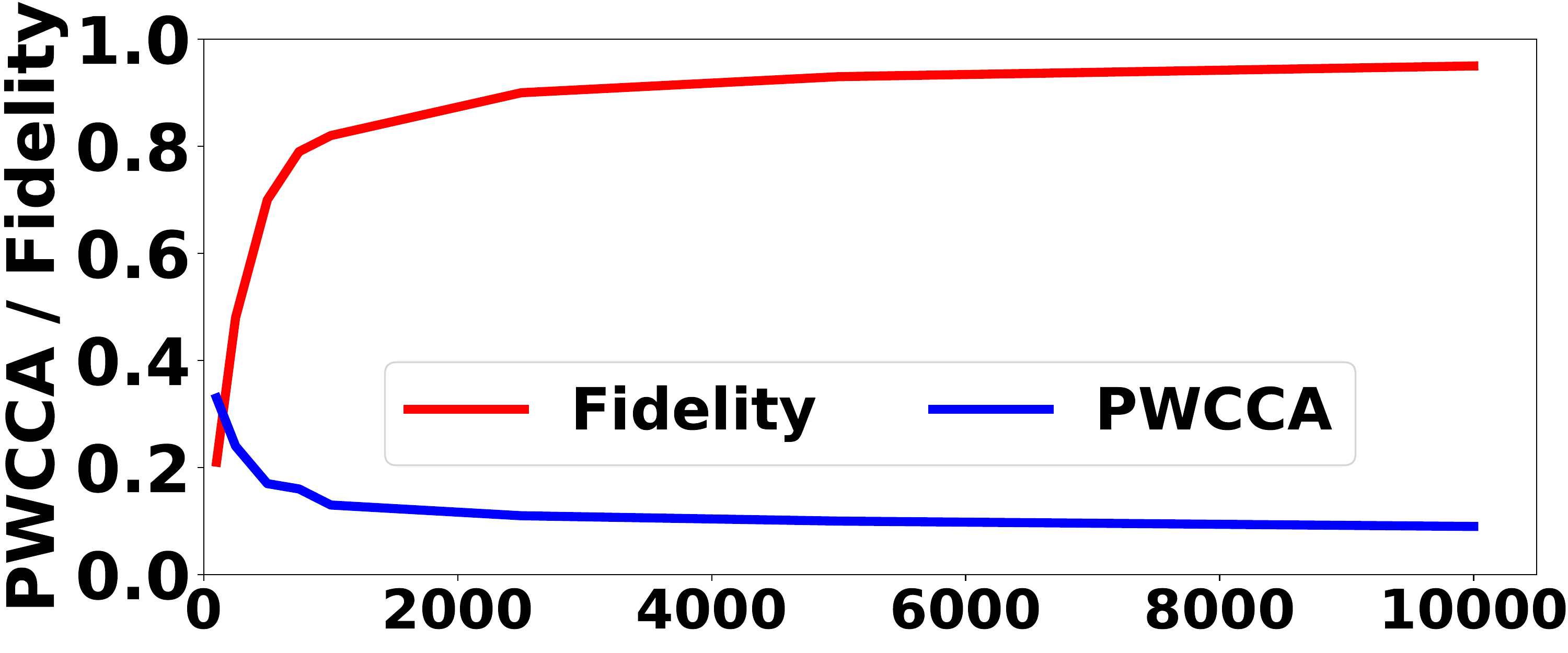}
    \caption{MNIST PWCCA vs Fidelity} \label{fig:pwcca_MNIST}
  \end{subfigure}%
  \hspace{0.02pt}
  \begin{subfigure}{0.23\textwidth}
    \includegraphics[width=\linewidth]{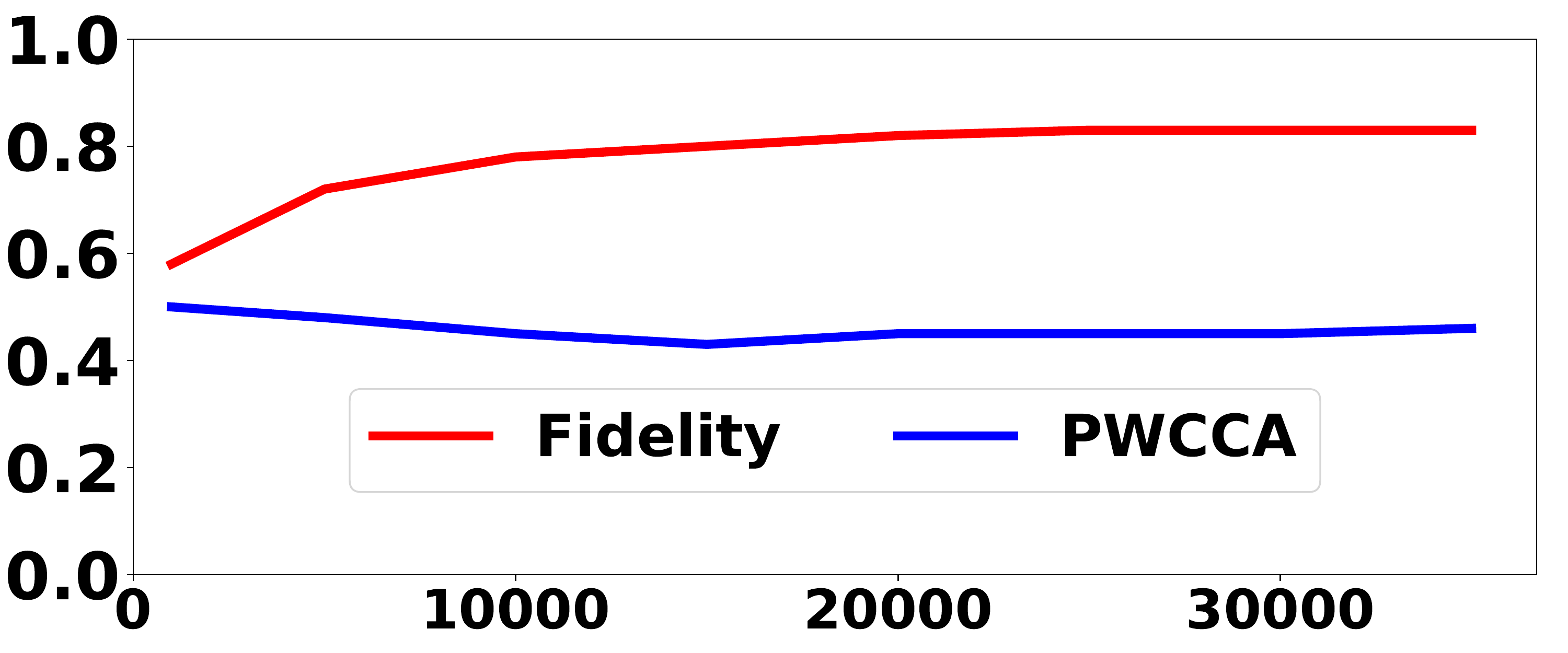}
    \caption{CelebA PWCCA vs Fidelity} \label{fig:pwcca_faces}
  \end{subfigure}%
  \hspace{0.02pt}
  \\
  \begin{subfigure}{0.48\textwidth}
    \includegraphics[width=\linewidth]{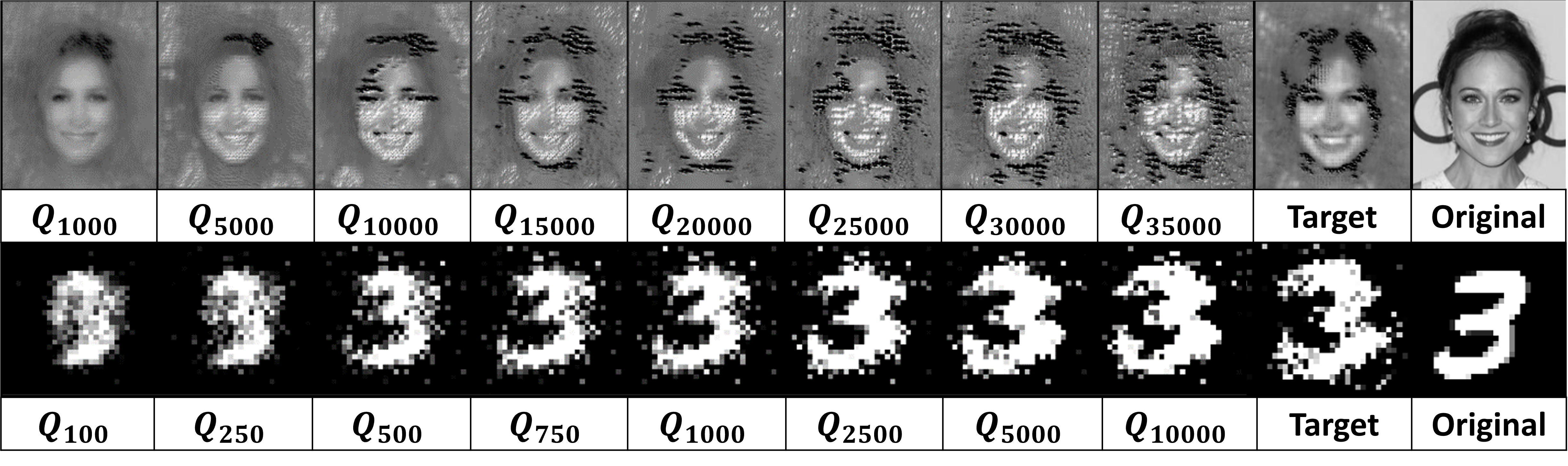}
    \caption{Inversion Results (CelebA (Top), MNIST (Bottom))} \label{fig:modelInversionResults}
  \end{subfigure}%
  \hspace{0.02pt}

 \caption{\label{fig:modelInversion} \textbf{Model inversion upon stolen models.} Using stolen models from KON across varying query amounts \textit{($Q_{n}$)} evaluated against target model, and original image.}
\end{figure}
%% %---------------------------

Extraction attacks have been identified as a means to stage further adversarial attacks \cite{deepSniffer, zhu2020hermes, surveyWANG201912}. Using PINCH, we conduct a case study whereby an adversary uses an extraction attack to stage a further model inversion attack.

%Two target models were stolen; Custom\_MNIST, and Custom\_CelebA trained on two datasets of differing complexity (MNIST 28x28, CelebA 218x178, both greyscale).
\textbf{Scenario setup}. Two target models; a 3-layer and 4-layer model architecture for MNIST and CelebA, respectively, were stolen by KON. The subsequent shadow model is then exposed to a model inversion attack \textit{MiFace}~\cite{MiFace}, whereby model information is used to generate images representative of target model classes. The MiFace attack was performed on the shadow model at various query requests for MNIST [100 -- 10000] and CelebA [1000 -- 35000] to evaluate MiFace attack success at different stolen model fidelity. Following the threat model in Section \ref{sec:threatmodel} the adversary has access to an auxiliary dataset, and partial knowledge of the target model. Thus images representative of model classes; written numbers for MNIST and random faces for CelebA, are used for image initialization \cite{modelInversion, MiFace}. MiFace attack success when applied to a stolen model was evaluated by comparing generated images against images generated by the target model.

\textbf{Inversion success}. Observing results from Figure \ref{fig:modelInversion}, it is apparent that even partial extraction success (0.83, 0.95) can result in successful model inversion exhibiting similar generated features comparable to the target. Specially, class features such as shape are captured to a high degree of accuracy. For example, CelebA successfully captured dataset specific features face shape, eye and mouth positioning inline with the actual image of the class. These captured features enable the adversary to gain additional knowledge about the target models, and therefore exploit this knowledge to further augment the model inversion attack, or repeat extraction with a more tuned query set. Dataset complexity also affects inversion success. A less complex dataset such as MNIST shows highly similar images, whereas CelebA -- a more complex dataset -- introduces more noise due to more granular greyscale and the image containing more distinctive features \cite{mnist, CelebA}.

\textbf{Extraction sensitivity}. Additionally, we demonstrate MiFace upon stolen models of varying query amounts and highlight that successful inversion occurs with low query amounts and fidelity varying on dataset complexity (Figure \ref{fig:modelInversionResults}). We observed that both datasets begin to show class features early with MNIST establishing a clear shape with 500 queries and similarly CelebA at 10,000 queries. Interestingly, we see that the fidelity of the CelebA stolen model shows signs of convergence with over 25,000 queries, however displays different model inversion results despite no increase in fidelity. 

\textbf{Architecture similarity}. To further investigate the similarity between target and stolen models, we applied Projection Weighted Canonical Correlation Analysis (PWCCA)\cite{CanCorrelation} that measures similarity by calculating the distance between the activation layers of two models during inference (Figure \ref{fig:pwcca_MNIST} and \ref{fig:pwcca_faces}). We observed that increasing query amounts for MNIST resulted in higher fidelity and PWCCA distance to decrease, denoting high similarity. CelebA also follows this trend until 15,000 queries where the PWCCA distance was at its lowest, however additional query amounts increase the PWCCA distance despite fidelity increasing. This is due to training being a high dimensional optimization problem, whereby their exists high numbers of \textit{Local Optimum} that can achieve similar fidelity and accuracy due to outputs have similar \cite{localoptimum}. These local optimum can have large parameter space between them, thus exhibiting a high PWCCA distance.

\section{Discussion}\label{sec:discussion}

%Characteristics Affecting Extraction Attack Success
\subsection{Key Extraction Characteristics}\label{sec:intrinsicResistance}

Within our evaluation in Section~\ref{sec:evaluation} we have identified multiple characteristics influencing attack success, each exhibiting a particular affinity for specific attacks (e.g. dataset complexity for KON, hardware for DR). We determined that differences within model architecture directly altered success of all attacks. As such, it is apparent that DL models appear to exhibit different resistance against certain attacks based on their model architecture characteristics. We suspect this phenomena also exists for other types of adversarial attacks. Further study of this finding would allow researchers to more effectively study and explain commonalities between adversarial attacks within complex networked systems and provide practitioners the ability to focus engineering effort (validation, countermeasures, design) to secure DL models against attacks with the highest likelihood of success based on their architectural characteristics.

\subsection{Further Attack Staging}

From conducting experiments, we observed that even with stolen models acquired from partially successful extraction attacks can be leveraged to stage further adversarial attacks such as model inversion to attain reasonable levels of success (0.7+ fidelity). Less complex datasets such as MNIST are considerably less noisy compared to more complex datasets such as CelebA. However defined features can still be extracted by the inversion attack such as face shape, and gender (Figure \ref{fig:modelInversion}). The ability to extract such features is especially concerning given the privacy related issues associated with specific types of models such as facial recognition, which allow an adversary to reverse engineer the classes to generate and expose images associated with real world people \cite{modelInversion, MiFace}. This highlights that underlying hidden knowledge present within DL models can be extracted from stolen models, and therefore adversarial attack staging must be studied further.

\begin{figure}[t]
  \includegraphics[width=\linewidth]{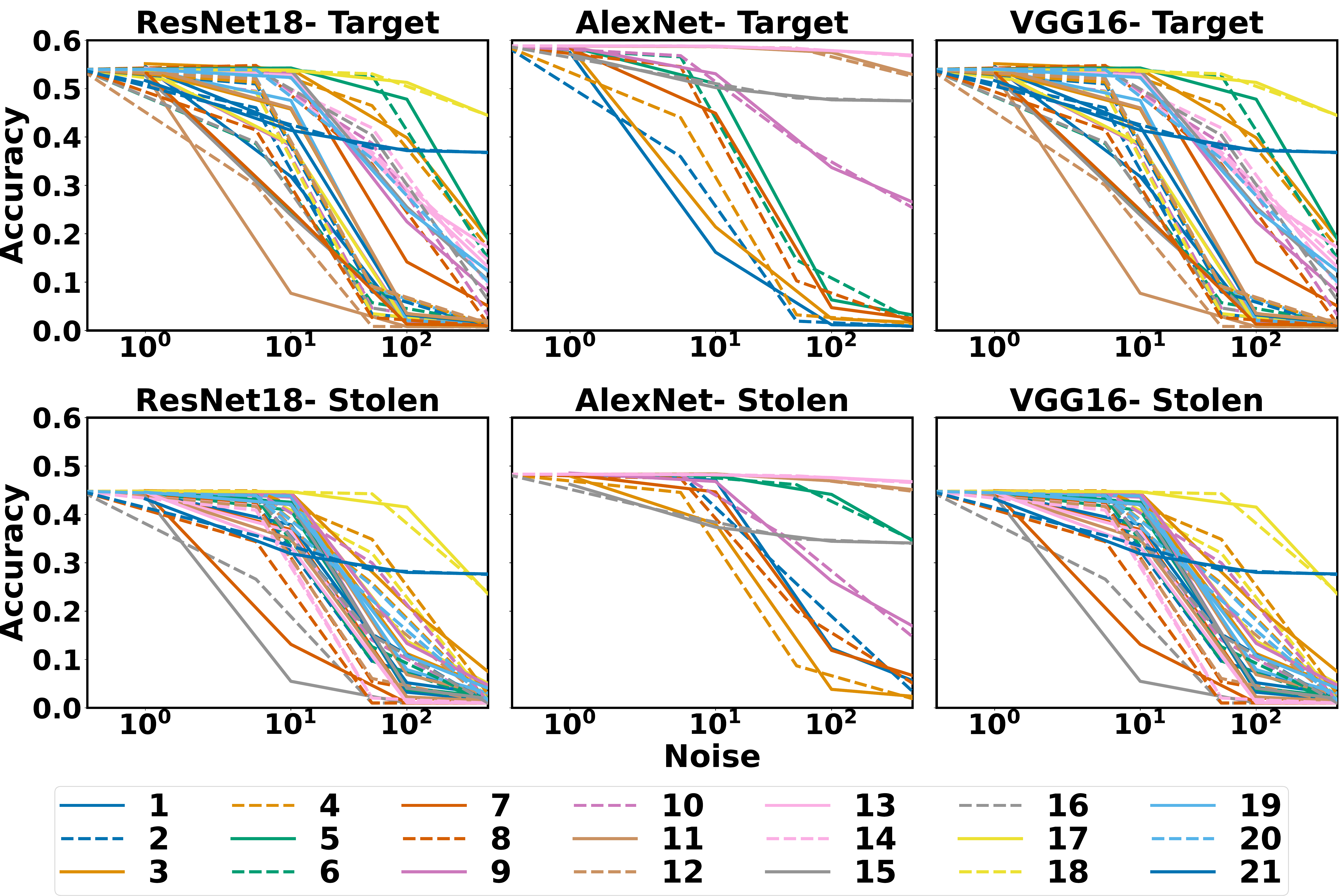}
  \hspace{0.04\textwidth}
  \caption{\label{fig:noiseResults} \textbf{Layer sensitivity to noise.} Each line represents the sensitivity of a CIFAR100 model layer to increasing magnitudes of noise, highlighting its expressive power.}
\end{figure}

\subsection{Extraction Equivalency} \label{sec:equivalency}

\textbf{Fidelity vs PWCCA.} We observe that due to the high dimensional optimization problem existing in training networks, Fidelity and PWCCA can contrast between each other due to the existence of many local optimum (See Section \ref{sec::furtherAttacks}). Despite not being exactly equivalent to their targets, stolen models still contain concerning information originating from the target model. This implicates that a new measurement of similarity is required to better understand the success of adversarial attacks, and further the creation of countermeasures. 
%Training neural networks is a high dimensional optimisation problem. There are many solutions to this problem, therefore there exists many optimal parameters of neural networks known as \textit{Saddle Points}. All saddle points can achieve the same fidelity because their outputs are the same image and can achieve the same model accuracy. Our training result, chose randomly, one of the potential saddle points. However the distance between two saddle points can be very large in parameter space. Thus, using PWCCA; a measurement of distance in parameters space, a model of high fidelity can exhibit a low PWCCA distance and therefore why correlation between PWCCA and fidelity does not always exist.

\begin{figure}[t]
  \begin{subfigure}{0.48\textwidth}
    \includegraphics[width=\linewidth]{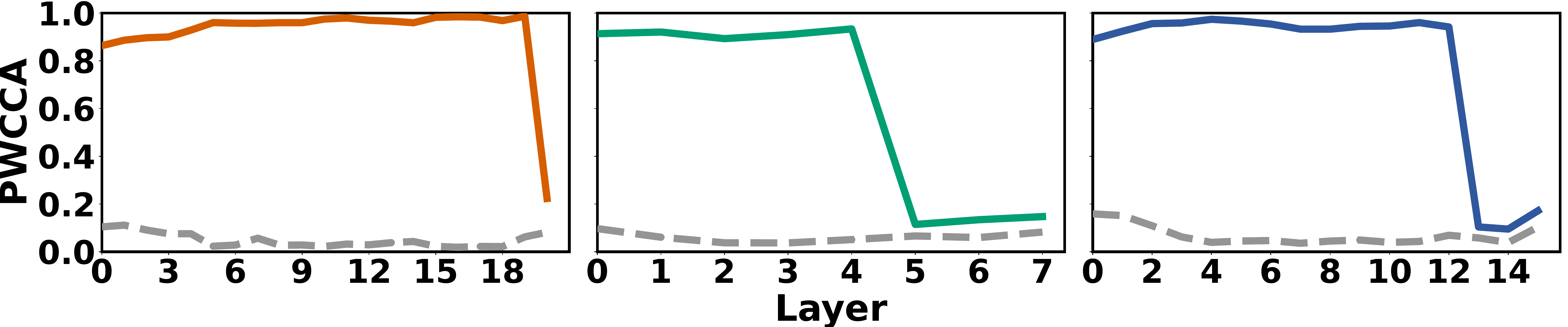}
    \caption{Original Models} \label{fig:original_models_dist}
  \end{subfigure}%
  \\
  \begin{subfigure}{0.48\textwidth}
    \includegraphics[width=\linewidth]{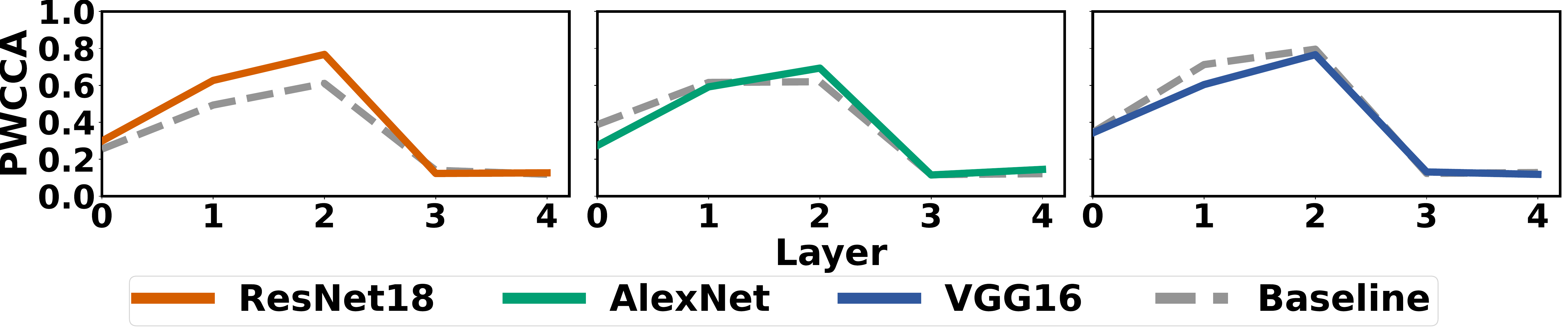}
    \caption{Distilled Models} \label{fig:stolen_models_dist}
  \end{subfigure}%
\caption{\label{fig:distillation} \textbf{Knowledge distillation upon stolen models.} PWCCA distance comparison of target and stolen models across ResNet18, AlexNet, and VGG16 further distilled into a 5 layer CNN \cite{distillation}. Baselines computed by comparing target models with an identically trained model.}
\end{figure}

% \begin{figure}[t]
% \includegraphics[width=\linewidth]{graphics/graph_pwcca_baselines_distillation.pdf}
% \caption{\label{fig:distillation} \textbf{Knowledge Distillation Upon Stolen Models} Target and stolen models of ResNet18, AlexNet, and VGG16 distilled into a 5 layer CNN compared with PWCCA distance. Baselines computed by comparing target models with an identically trained model.}
% \end{figure}

\textbf{Expressive power of models.} We analyzed the expressiveness of the target and stolen model to investigate the layer sensitivity to noise by utilizing the method proposed in \cite{expressivePower}. We compared the sensitivity of the target and stolen model to increasing magnitudes of noise to explore if the same layers of these two models exhibit the same pattern in accuracy degradation (Figure \ref{fig:noiseResults}). Similarly to the findings within the proposed method, stolen models hold true to the statement \textit{trained networks are more sensitive to their lower (initial) layer weights}, as earlier layers in stolen models were also most sensitive. Additionally, target and stolen models exhibit different sensitivity to noise across the majority of layers within the network, highlighting how fundamentally the models have learned differently to each other. Further understanding the difference in expressiveness would benefit the development of adversarial defences and secure DL models by providing foresight into how stolen models adapt target model knowledge into their own expressive structure.

%Target models in Figure \ref{fig:target_models_noise} show different most sensitive layers in different models when compared to their stolen counterpart in Figure \ref{fig:stolen_models_noise}, despite achieving relatively good accuracy comparable to the target. 

%A method to analyse expressiveness of DNN's was explored by Raghu \textit{et al.} whereby various magnitudes of noise was applied to each layer within a network to observe which layers are most sensitive to noise and therefore are critical to their accuracy \cite{expressivePower}. Using this method, we compare how a target and stolen model differ in their reaction to noise, therefore understanding if the same layers exhibit the same pattern, and thus are similar. We observe that stolen models follow the same assumption made by Raghu \textit{et al.} where; \textit{Trained networks are more sensitive to their lower (initial) layer weights}, as earlier layers in stolen models were also most sensitive (Figure \ref{fig:noiseResults}). 

%Knowledge Distillation aims to produce a smaller model by taking the most critical features of a teacher model. Therefore creating a distilled model of the target and stolen models we can evaluate if the stolen model was able to capture the same critical features stored within the target model. 
\textbf{Knowledge capture} We explored how much similar knowledge a stolen model captured from a target model. Knowledge Distillation (KD) \cite{distillation} was used to transfer the extracted knowledge of the target and stolen models into small distilled models, and then PWCCA is exploited to measure the representation similarities of these distilled models. 
Our experimental results indicate that similar captured knowledge exists between the target and stolen models. As shown in Figure \ref{fig:distillation}, we distilled the target and stolen model of 3 DNNs into smaller 5 layer CNNs, achieving accuracy's stated in Table \ref{appendix:distillationStats}. The PWCCA distance of the distilled models in Figure \ref{fig:stolen_models_dist} is drastically lower compared to the original models in Figure \ref{fig:original_models_dist}, indicating that stolen models are capable of capturing knowledge contained within a target model despite being expressively different (Figure \ref{fig:noiseResults}).

\section{Related Work} \label{sec:relatedwork}
There is a growing body of research dedicated towards the study of adversarial attacks against DL model architectures and datasets within DL systems \cite{akhtar2018threat} \cite{chakraborty2018adversarial} \cite{chakraborty2021survey} \cite{huang2017adversarial} \cite{mani2021defending} \cite{qiu2020adversarial} \cite{ren2020adversarial}. 

\textbf{Extraction attack studies}. Tram\`{e}r \textit{et al.} \cite{stealingPredictionAPIs} introduced the first extraction attack to extract target ML models exposed in online prediction APIs. Papernot \textit{et al.} \cite{papernot2017practical} proposed an avatar approach to extract a substitute DNN model for the purpose of generating adversarial examples. Different from \cite{papernot2017practical}, Joon \textit{et al.} \cite{oh2019towards} designed an avatar based approach to train a meta-model to predict model hyperparameters. Junti \textit{et al.} \cite{juuti2019prada} developed a generic method for extracting DNN models by optimizing training hyperparameters and generating synthetic queries. Orekondy \textit{et al.} \cite{knockOffDLs} proposed a reinforcement learning based framework to improve query sample efficiency and performance. Hua \textit{et al.} \cite{hua2018reverse} first studied on reverse engineering of CNN on hardware accelerators, and investigated potential vulnerabilities in CNN accelerators in the context of model stealing. Wang \textit{et al.} \cite{wang2018stealing} provided hyperparameter stealing attacks to DL models.
 
\textbf{Adversarial attack frameworks}.
Hussain \textit{et al.} \cite{PrivacyRaven} presented a library allowing for black-box and label-only extraction, inference and inversion attacks on DL models. As an extended work of \cite{PrivacyRaven}, Nicolae \textit{et al.} \cite{Nicolae2018AdversarialRT} developed a more feature-rich library for evaluating and defending ML models to extraction, inference, inversion and poisoning attacks. Chen \textit{et al.} \cite{chen2020frank} designed a Frank-Wolfe algorithm-based adversarial attack framework for white-box and black-box settings. Pearce \textit{et al.} \cite{Counterfit} provided a generic automation tool for testing the security of ML.
% , which is a flexible environment, model and data agnostic framework
Liu \textit{et al.} \cite{mldoctor} proposed a holistic risk assessment of different inference attacks against ML models and established a threat model taxonomy. In contrast to these works, we propose an end-to-end automated extraction attack framework capable of conducting an in-depth evaluation of extraction attacks across various operational scenarios and heterogeneous hardware platforms.

%-------------------------------------------------------------------------------
\section{Conclusion}\label{sec:conclusion}
%In this paper we have proposed a framework for studying adversarial extraction attacks. The framework is capable of rapidly designing, deploying, and analyzing a large number of extraction attack scenarios across a multitude of different DL model architectures, datasets, hardware, and attack types. Results identify key characteristic influencing attack success, demonstrate extraction can achieve as high as 95\% similarity, and show that even partial success can enable further privacy concerning adversarial attacks. Additionally we highlight the existence of intrinsic susceptibility whereby specific model configurations exhibited strong resilience to attacks Equivalency of these stolen models via expressiveness etc to their targets.

In this paper we have conducted an extensive empirical experimentation of extraction attack scenarios. Utilizing PINCH to rapidly design, deploy, and analyze a large number of extraction attack scenarios not yet captured in current literature, we have conducted a detailed study of extraction effectiveness against different DL system environments. We identify key insights into the fundamental understanding of adversarial security: We have (1) uncovered key extraction characteristics whereby specific model configurations exhibit strong resilience to specific attacks; (2) stolen models exhibit equivalent functionality with fundamentally different model characteristics and expressive power; and (3) demonstrated even partial extraction success enables staging of further privacy concerning adversarial attacks.

%-------------------------------------------------------------------------------
% \section*{Acknowledgment}
% This work was supported by the EPSRC (EP/V026763/1, EP/V007092/1).

\bibliographystyle{plain}
\bibliography{bibfile}

\clearpage
\onecolumn
\appendix

\counterwithin{table}{section}

\section{Appendix: Additional Experimental Results}
In this appendix, we report additional results complimentary to experiments mentioned throughout the paper.

\setcounter{table}{0}
\begin{table}[htbp]
\begin{center}
\resizebox{\columnwidth}{!}{%
\begin{tabular}{cccllllllllc}
\hline
\multirow{2}{*}{\textbf{System}}                                              & \multirow{2}{*}{\textbf{\begin{tabular}[c]{@{}c@{}}Cache\\ (MB)\end{tabular}}} & \multirow{2}{*}{\textbf{}} & \multicolumn{8}{c}{\textbf{Feature Prevalence}}                                                                                                                                                                                                                                                     & \multirow{2}{*}{\textbf{Simplified}}        \\
                                                                              &                                                                                &                            & \multicolumn{1}{c}{\textbf{convs}} & \multicolumn{1}{c}{\textbf{fcs}} & \multicolumn{1}{c}{\textbf{softms}} & \multicolumn{1}{c}{\textbf{relus}} & \multicolumn{1}{c}{\textbf{mpool}} & \multicolumn{1}{c}{\textbf{apool}} & \multicolumn{1}{c}{\textbf{merge}} & \multicolumn{1}{c}{\textbf{bias}} &                                             \\ \hline
\multirow{2}{*}{\textbf{\begin{tabular}[c]{@{}c@{}}i5-3470\end{tabular}}} & \multirow{2}{*}{4}                                                             & X                          & 0.0003                             & -0.1643                          & -0.0583                             & 0.6807                             & -0.6021                            & 0.3475                             & -0.069                             & -0.1340                           & relus\textgreater{}mpool\textgreater{}apool \\
                                                                              &                                                                                & Y                          & -0.0004                            & -0.4484                          & 0.3590                              & -0.4292                            & -0.0839                            & 0.5101                             & 0.373                              & -0.2805                           & apool\textgreater{}relus\textgreater{}fcs   \\ \hline
\multirow{2}{*}{\textbf{\begin{tabular}[c]{@{}c@{}}i7-4770\end{tabular}}} & \multirow{2}{*}{8}                                                             & X                          & -0.8412                            & 0.0808                           & 0.0776                              & 0.0119                             & 0.0625                             & 0.1082                             & 0.5136                             & -0.0136                           & convs\textgreater{}merge\textgreater{}apool \\
                                                                              &                                                                                & Y                          & 0.3603                             & -0.0684                          & -0.0680                             & -0.6455                            & -0.0605                            & -0.0983                            & 0.6524                             & -0.0718                           & merge\textgreater{}relus\textgreater{}convs \\ \hline
\multirow{2}{*}{\textbf{\begin{tabular}[c]{@{}c@{}}i7-6850k\end{tabular}}} & \multirow{2}{*}{15}                                                            & X                          & -0.7878                            & -0.0029                          & -0.0038                             & 0.5642                             & -0.0054                            & 0.0044                             & 0.2462                             & -0.0148                           & convs\textgreater{}relus\textgreater{}merge \\
                                                                              &                                                                                & Y                          & -0.1529                            & 0.0051                           & 0.0001                              & -0.5671                            & -0.0118                            & -0.0081                            & 0.8060                             & -0.0712                           & merge\textgreater{}relus\textgreater{}convs \\ \hline
\end{tabular}%
}
\caption{\textbf{Hardware platforms evaluated against the DeepRecon side-channel attack, and PCA feature prevalence.} Well-fingerprinted systems i7-4770 and i7-6850k closely align on the most prevalent features (\textit{Conv}, \textit{Merge}, \textit{ReLU}).} \label{appendix:deepReconTestSuite}
\end{center}
\end{table}

\begin{table}[]
\begin{center}
\begin{tabular}{ccccccc}
\hline
\textbf{Name}    & \textbf{Family}            & \textbf{Parameters} & \multicolumn{1}{l}{\textbf{GMAdd}} & \textbf{KON} & \textbf{DS} & \textbf{DR} \\ \hline
Alexnet          & -                          & 61.10m              & 0.72                               & Y            & Y           & Y           \\
ConvNeXt\_Small  & \multirow{2}{*}{ConvNeXt}  & 50.21m              & 8.70                               & Y            & Y           & Y           \\
ConvNeXt\_Large  &                            & 197.74m             & 34.40                              & Y            & Y           & Y           \\
Densenet121      & \multirow{4}{*}{Densenet}  & 7.99m               & 2.88                               & Y            & Y           & Y           \\
Densenet161      &                            & 28.68m              & 7.82                               & Y            & Y           & Y           \\
Densenet169      &                            & 14.14m              & 3.42                               & Y            & Y           & N           \\
Densenet201      &                            & 20.01m              & 4.37                               & N            & N           & Y           \\
Resnet18         & \multirow{5}{*}{Resnet}    & 11.68m              & 1.82                               & Y            & Y           & N           \\
Resnet34         &                            & 21.79m              & 3.68                               & Y            & Y           & N           \\
Resnet50         &                            & 25.55m              & 4.12                               & Y            & Y           & N           \\
Resnet101        &                            & 44.55m              & 7.85                               & N            & N           & Y           \\
Resnet152        &                            & 60.19m              & 11.58                              & N            & N           & Y           \\
VGG11            & \multirow{4}{*}{VGG}       & 132.86m             & 7.63                               & Y            & Y           & N           \\
VGG13            &                            & 133.04m             & 11.34                              & Y            & Y           & N           \\
VGG16            &                            & 138.358m            & 15.50                              & Y            & Y           & Y           \\
VGG19            &                            & 143.67m             & 19.67                              & N            & N           & Y           \\
RegNetY-400MF    & -                          & 4.344m              & 0.41                               & Y            & Y           & N           \\
SqueezeNet       & -                          & 1.248m              & 0.83                               & Y            & Y           & N           \\
ViTB16           & -                          & 86.568m             & 11.28                              & Y            & Y           & Y           \\
MobileNetV2      & -                          & 3.50m               & 0.32                               & Y            & Y           & Y           \\
InceptionV3      & - & 27.161m             & 2.85                               & N            & N           & Y           \\ \hline                    
\end{tabular}
\caption{ {\upshape \textbf{Experiment model architectures.} (KON=KnockOffNets, DS=DeepSniffer, DR=DeepRecon).}} \label{appendix:targetArchitectures}
\end{center}
\end{table}

% \usepackage{multirow}
% \usepackage{graphicx}
\begin{table}[]
\centering
\resizebox{\textwidth}{!}{%
\begin{tabular}{llllllll}
\hline
\multicolumn{1}{c}{\multirow{2}{*}{\textbf{Model}}} & \multicolumn{1}{c}{\multirow{2}{*}{\textbf{\begin{tabular}[c]{@{}c@{}}Average Exact Accuracy\\ Across All Hardware\\ (1 d.p.)\end{tabular}}}} & \multicolumn{1}{c}{\multirow{2}{*}{\textbf{Family}}} & \multicolumn{1}{c}{\multirow{2}{*}{\textbf{\begin{tabular}[c]{@{}c@{}}Average Family Accuracy\\ Across All Hardware\\ (1 d.p.)\end{tabular}}}} & \multicolumn{4}{c}{\textbf{\begin{tabular}[c]{@{}c@{}}Hardware Average\\ Family Accuracy\\ (1 d.p.)\end{tabular}}} \\
\multicolumn{1}{c}{} & \multicolumn{1}{c}{} & \multicolumn{1}{c}{} & \multicolumn{1}{c}{} & \textbf{i5-3470} & \textbf{i7-3770k} & \textbf{i7-4770} & \textbf{i7-6850k} \\
 &  &  &  &  &  &  &  \\ \hline
AlexNet & 0.50 & AlexNet & 0.50 & 0.29 & 0.19 & \textbf{0.84} & 0.67 \\ \hline
DenseNet121 & 0.52 & \multirow{3}{*}{DenseNet} & \multirow{3}{*}{0.84} & \multirow{3}{*}{0.52} & \multirow{3}{*}{0.89} & \multirow{3}{*}{0.93} & \multirow{3}{*}{\textbf{1.0}} \\
DenseNet161 & 0.31 &  &  &  &  &  &  \\
DenseNet201 & 0.43 &  &  &  &  &  &  \\ \hline
MobileNetV2 & 0.35 & MobileNet & 0.35 & 0.13 & 0.09 & 0.37 & \textbf{0.83} \\ \hline
ResNet50 & 0.42 & \multirow{3}{*}{ResNet} & \multirow{3}{*}{0.43} & \multirow{3}{*}{0.21} & \multirow{3}{*}{0.40} & \multirow{3}{*}{0.35} & \multirow{3}{*}{\textbf{0.77}} \\
ResNet101 & 0.52 &  &  &  &  &  &  \\
ResNet152 & 0.47 &  &  &  &  &  &  \\ \hline
VGG16 & 0.39 & \multirow{2}{*}{VGG} & \multirow{2}{*}{0.77} & \multirow{2}{*}{0.43} & \multirow{2}{*}{0.91} & \multirow{2}{*}{0.81} & \multirow{2}{*}{\textbf{0.91}} \\
VGG19 & 0.41 &  &  &  &  &  &  \\ \hline
\end{tabular}%
}
\caption{\textbf{DeepRecon Average Results Across Models.} Includes results collected across all runs and hardware.}
\label{tab:deeprecon_averages}
\end{table}

\begin{figure}[h]
  \includegraphics[width=\linewidth]{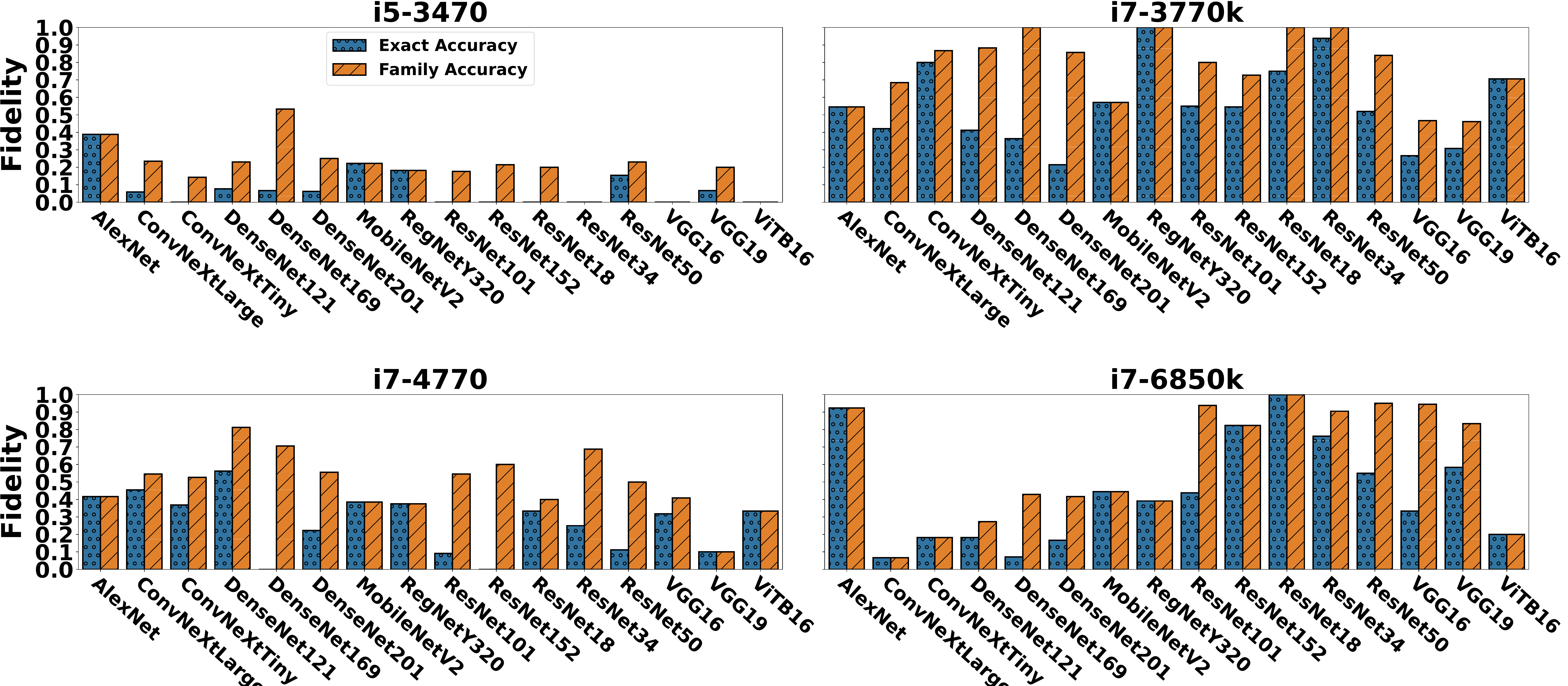}
\caption{\label{fig:deeprecon_accuracies_tf2} \textbf{DeepRecon model architecture \& family prediction for TensorFlow 2.10.}}
\end{figure}

\begin{table}[]
\begin{center}
\begin{tabular}{ccc}
\hline
\textbf{Name} & \textbf{PyTorch} & \textbf{TensorFlow} \\  \hline
VGG16         & 0.64             & -28.76              \\
MobileNet\_V2 & 0.41             & -3.76               \\
ResNet50      & 0.57             & -6.96               \\
Inception\_V3 & 0.13             & -10.56 \\ \hline               
\end{tabular}
\caption{ {\upshape \textbf{DeepSniffer Framework Comparison PyTorch vs TensorFlow.} Recorded fidelity between actual and predicted architecture from PyTorch and TensorFlow frameworks. TensorFlow returns unusable results.}} \label{appendix:pytorchVTensorflow}
\end{center}
\end{table}

\begin{table}[]
\begin{center}
\begin{tabular}{ccccc}
\hline
\textbf{Name} & \textbf{Target} & \textbf{Stolen} & \textbf{Distilled Target} & \textbf{Distilled Stolen} \\  \hline
ResNet18      & 0.54            & 0.45            & 0.35                      & 0.36                      \\
AlexNet       & 0.59            & 0.48            & 0.36                      & 0.35                      \\
VGG16         & 0.61            & 0.44            & 0.36                      & 0.35 \\ \hline                     
\end{tabular}
\caption{ {\upshape \textbf{Distillation Accuracy Results.} Accuracy scores based on CIFAR100 test set upon original and distilled target and stolen models.}} \label{appendix:distillationStats}
\end{center}
\end{table}

%%%%%%%%%%%%%%%%%%%%%%%%%%%%%%%%%%%%%%%%%%%%%%%%%%%%%%%%%%%%%%%%%%%%%%%%%%%%%%%%
\end{document}